\documentclass[journal]{IEEEtran}
\usepackage{array}
\usepackage{amsmath}
\usepackage{fixltx2e}
\usepackage{stfloats}
\usepackage{cite}
\usepackage[pdftex]{graphicx}
\usepackage{subfigure}
\usepackage{graphicx}
\usepackage{booktabs}
\usepackage{graphicx}
\usepackage{float}
\usepackage{caption}
\usepackage{times,amsmath,epsfig}
\usepackage[numbers,sort&compress]{natbib}
\usepackage{epstopdf}
\usepackage{cleveref}
\usepackage{multirow}
\usepackage{algorithm}  
\usepackage{algpseudocode}  
\usepackage{url}

\begin{document}
%
\title{Statistical Learning Based Congestion Control \\for Real-time Video Communication}
%
%
%
%

\author{Tongyu~Dai,
        Xinggong~Zhang,
        Yihang~Zhang,
        and~Zongming~Guo,~\IEEEmembership{Member,~IEEE}
        }

\maketitle

\begin{abstract}

With the increasing demands on interactive video applications, how to adapt video bit rate to avoid network congestion has become critical, since congestion results in self-inflicted delay and packet loss which deteriorate the quality of real-time video service. The existing congestion control is hard to simultaneously achieve low latency, high throughput, good adaptability and fair bandwidth allocation, mainly because of the hardwired control strategy and egocentric convergence objective.

To address these issues, we propose an end-to-end statistical learning based congestion control, named Iris. By exploring the underlying principles of self-inflicted delay, we reveal that congestion delay is determined by sending rate, receiving rate and network status, which inspires us to control video bit rate using a statistical-learning congestion control model. The key idea of Iris is to force all flows to converge to the same queue load, and adjust the bit rate by the model. All flows keep a small and fixed number of packets queuing in the network, thus the fair bandwidth allocation and low latency are both achieved. Besides, the adjustment step size of sending rate is updated by online learning, to better adapt to dynamically changing networks.


We carried out extensive experiments to evaluate the performance of Iris, with the implementations of transport layer (UDP) and application layer (QUIC) respectively. The testing environment includes emulated network, real-world Internet and commercial LTE networks. Compared against TCP flavors and state-of-the-art protocols, Iris is able to achieve high bandwidth utilization, low latency and good fairness concurrently. Especially over QUIC, Iris is able to increase the video bitrate up to 25\%, and PSNR up to 1dB.

\end{abstract}


%
\IEEEpeerreviewmaketitle

\section{Introduction}
\label{sec:intro}

\IEEEPARstart{W}{ith} the widespread deployments of LTE/WiFi wireless networks and the forthcoming 5G \cite{5G}, interactive video applications are growing exponentially, from the mobile video chatting, such as Skype \cite{skype}, FaceTime, to AR/VR streaming \cite{youtube_vr} and cloud gaming \cite{cloud_gaming,cloudgaming}. These video applications require not only higher bandwidth but also lower transmission delay. However, real-world network capacity is constrained, especially in wireless networks with unpredictable dynamics (e.g. random packet loss, channel fading, etc) \cite{LTE_study,wireless_effect}. It imposes great challenges on nowadays video bitrate adaptation, which adjusts video streaming bit rate to reduce self-inflicted delay and  loss.

The existing work related to rate adaptation can be mainly divided into two categories. The first is the research of adaptive bit rate (ABR) algorithms over application layer, including \cite{mdash,XMAS,pensieve,tmm18_dash}. They usually use HTTP as the transport protocol and adjust the video bit rate according to bandwidth estimation and buffer state \cite{XMAS}. However, the underlying TCP of HTTP will essentially bring high latency \cite{bufferbloat}, which is not suitable for real-time video transmission.

\begin{figure}[t]
\centering
\includegraphics[width=3.4in]{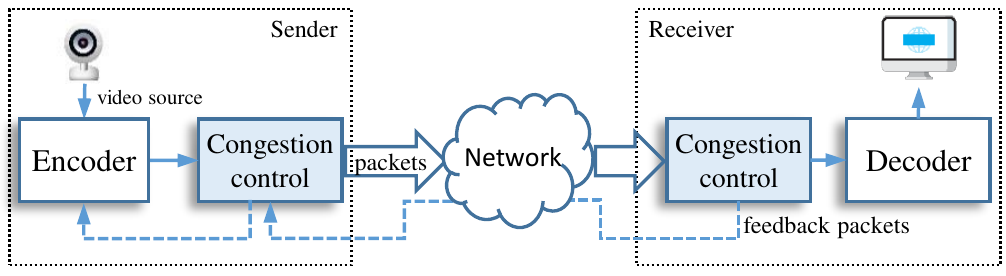}
\caption{The architecture of real-time video transmission.}
\label{architecture}
\end{figure}

The second kind of rate adaptation related work is congestion control based on transport layer. As shown in \Cref{architecture}, congestion control plays an important role in the real-time video transmission, which adapts video bit rate to avoid self-inflicted delay and packet loss. Many rate adaptation schemes \cite{WebRTC_coding,multi_tcp,tmm12_ll,tmm12_MTCC,tmm17_agent,tmm16_jihe} have been proposed for video calling or video conferencing. Most of them focus on rate allocation upon given bandwidth \cite{WebRTC_coding,multi_tcp} or relay server selection \cite{tmm17_agent,tmm16_jihe}.  Seldom of them considers how to reduce end-to-end congestion delay. In addition to the conventional TCP-like algorithms \cite{cubic,tcp-vegas}, there are also some end-to-end congestion control designed for real-time video streaming or low-latency transmission, which can limit the self-inflicted delay \cite{gcc,tmm12_ll,2016BBR} or achieve TCP fairness \cite{tmm12_MTCC}. Some methods also adjust video bit rate with online learning \cite{2018pcc,online_tsp}. But these existing algorithms still have some flaws, mainly including the following two aspects.






\begin{itemize}

  \item \textbf{Non-coexistence of high throughput and low latency}. Considering only packet loss will lead to high queuing delay, but the algorithms overly concerned about delay also lead to low throughput \cite{vegas-overestimate}.
  

  \item \textbf{Hardwired control strategy}. Most methods  adjust sending rate with fixed step size or multiplier \cite{TCP-friendly}. This manual mapping cannot always be optimal in changing networks, resulting in performance degradation.
    
  \item  \textbf{Egocentric convergence objective}. Some algorithms based on objective function are self-centered and lack communication between concurrent data streams, thus thye can not keep the same convergence goal for different clients, which leads to unfair bandwidth allocation.
  
\end{itemize}

Motivated by these issues, to obtain higher video transmission quality, we probe into the essence of congestion control and consider whether it is possible to design an algorithm that achieves the goals of low latency, high channel utilization, good adaptability to changing networks and fair bandwidth allocation. Some recently proposed algorithms \cite{Remy,remy_fault} have enlightened us: a congestion control with low-latency fairness objective and a learned rate adjustment strategy.

In this paper, we start with an in-depth investigation into network congestion in LTE, WiFi and Internet, to explore the underlying principles  of congestion delay. The data-driven analysis  reveals that there is a strong correlation between transmission latency, sending rate and receiving rate. Inspired by the observations, we have designed Iris, an end-to-end learning-based congestion control algorithm for real-time video streaming. It mainly consists of two components: \emph{low-latency fairness model} and \emph{learning-based rate control strategy}. The fairness model forces all streams to keep a small and fixed number of packets queuing in network, achieving fairness and low congestion delay. The learning-based rate control builds a statistical function between round-trip time (RTT), sending rate and receiving rate, based on online linear regression learning. Then it is used to determine the proper sending rate, which avoids fixed adjustment step size and converges to the fairness objective more quickly.

The contributions of this paper can be summed up in the following three aspects:

\begin{itemize}

  \item We explore the underlying reasons for congestion delay in video transmission and reveal the correlation between transmission latency, sending rate and receiving rate, which inspires us to design a low-latency algorithm with statistical learning.

  \item An adaptive bitrate adjustment scheme is introduced. According to the learning model, the rate adjustment step size can be periodically updated online, which avoids the hardwired control strategy to better adapt to dynamically changing networks.

  \item  A low-latency fairness model which can be proved theoretically is introduced. With the estimation of network status, the fairness model indicates the direction and size of delay adjustment, so that a fair share of bandwidth and low latency are guaranteed.
  
\end{itemize}

The paper is organized as follows: \Cref{Related Work} introduces the related work at first. \Cref{motivation} highlights the motivation of this paper. \Cref{algorithm}  describes our proposed Iris algorithm and the details of implementation are shown in \Cref{Implement}. \Cref{Evaluation} shows the experimental results and corresponding analysis.  \Cref{CONCLUSIONS} concludes the paper.

\section{Related Work}
\label{Related Work}

Conventional congestion control algorithms can be mainly divided into two categories: loss-based and delay-based. The loss-based methods, starting from Reno \cite{Reno} and extending to Cubic \cite{cubic} and Compound \cite{Compound}, interpret packet loss as the fundamental congestion signal. They continually push packets into the buffer of bottleneck link until packet loss occurs, resulting in ``bufferbloat'' and high queuing delay \cite{bufferbloat}. Besides, in wireless networks, low bandwidth utilization also results from the stochastic loss unrelated to congestion \cite{TCP_wireless}. To address these issues, the algorithms like Vegas \cite{tcp-vegas}, FAST \cite{tcp-fast} and LEDBAT \cite{ledbat} use delay, rather than packet loss, as the congestion signal. They perform well in constraining queuing delay, but overestimate delay because of ACK compression or network jitter, resulting in inadequate utilization of bandwidth \cite{vegas-overestimate}. Moreover, they will be starved when sharing a bottleneck link with concurrent loss-based flows \cite{vegas-starved}. Therefore, these conventional methods are not suitable for real-time video streaming.

There are also many studies focusing on special cases of network environments, including the algorithm customized for datacenters \cite{DCTCP,less,pfabric}, cellular networks \cite{sprout,adaptive}, Web applications \cite{webcc} and so on. These solutions yield good performance under the specific network conditions, but can not improve the performance of video streaming transmission. The algorithms specially designed for real-time video transmission mainly include \cite{gcc,nada,tmm12_ll}. They use packet loss and delay as congestion metrics, and empirically set some fixed thresholds. When the congestion metrics are higher or lower than these thresholds, the sending rate is adjusted accordingly. As we mentioned earlier, this hardwired control strategy can not maintain effectiveness in changing networks, resulting in performance degradation.


Over the past decade, more and more researchers have abandoned the TCP-like hardwired mapping rate control which maps events to fixed reactions (e.g. using fixed thresholds or step sizes). They prefer to generate effective control strategies through algorithms rather than handicraft, such as PCC \cite{2015PCC}, Verus \cite{2015Verus}, Remy \cite{Remy} and Vivace \cite{2018pcc}. Especially, Remy replaces the human designed algorithm with an offline optimization scheme that searches for the best scheme within its assumed network scenario. But the performance drastically degrades when the actual network conditions deviate from its assumptions \cite{remy_back}. PCC and Vivace propose to empirically observe and adopt actions that result in higher performance, but they have to try many times before deciding on the best action. The lag selection will affect the delay performance of real-time video streaming. Google proposes BBR \cite{2016BBR} to address the limitations of conventional TCP, which tries to make the number of inflight packets (i.e. data sent but not yet acknowledged) converge to bandwidth-delay product (BDP). However, it has poor performance in TCP-fairness, and its throughput will fluctuate periodically and violently due to its synchronization mechanism, which destroys the smoothness of video bitrate \cite{dai}.


In this paper, we propose a learning-based congestion control algorithm. Compared against these existing works, our innovations (and differences) lie in the following aspects:

\begin{itemize}

  \item \textbf{Convergence objective}. Although we need to obtain delay information in  Iris, its explicit convergence objective is the load of the queue, as described in \Cref{Model}. This is different from the traditional delay-based algorithms in which explicit delay values are used as congestion benchmarks. \emph{It can avoid the latecomer issue \cite{ledbat} of delay-based algorithms and improve fairness performance, but does not need to adopt hard synchronization like BBR \cite{2016BBR}}.
  
  \item \textbf{Environmental adaptation}. Iris introduced a rate adjustment method based on statistical learning, as described in \Cref{Rate Adjustment}, to balance the generality and specificity of the algorithm. Based on the collected historical data, Iris can adaptively change the rate adjustment step size without setting specific parameters for each network.
  
  \item  \textbf{Low overhead}. Different from the existing work that combines the concept of "learning", Iris uses a simple but effective linear regression learning model, which guarantees low overhead in complexity and time consumption. This approach avoids the drawbacks of Remy's offline learning due to its complexity \cite{remy_back}, and is better at real-time than PCC and Vivace.
  
\end{itemize}

\section{Data Study of Congested Network}
\label{motivation}  

\begin{figure}[t]  
\centering       
\subfigure[LAN.]{  
\begin{minipage}[t]{0.5\linewidth}
\centering  
\includegraphics[width=1.7in]{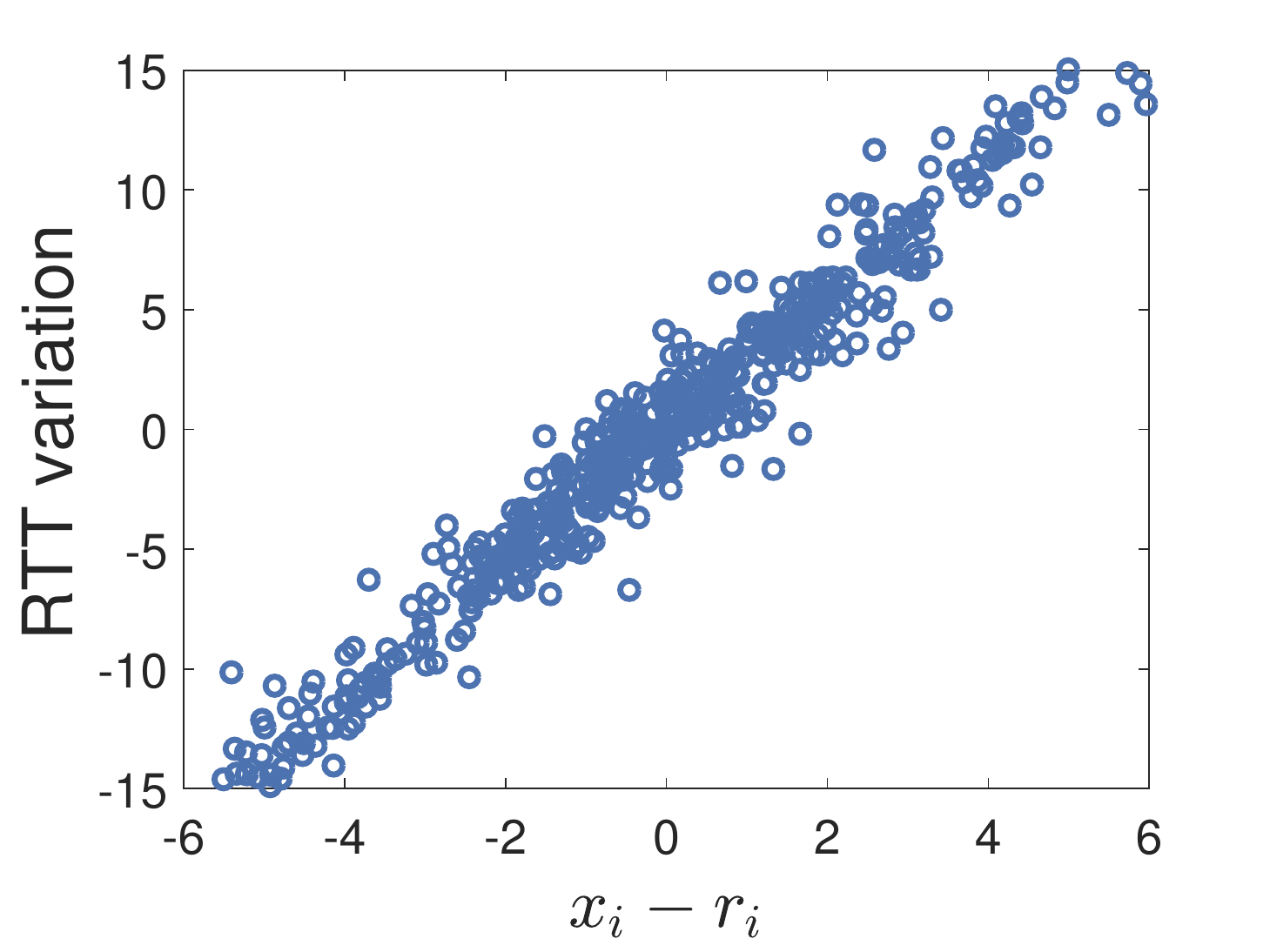}    
\end{minipage}%
}%
\subfigure[WAN.]{  
\begin{minipage}[t]{0.5\linewidth}  
\centering  
\includegraphics[width=1.7in]{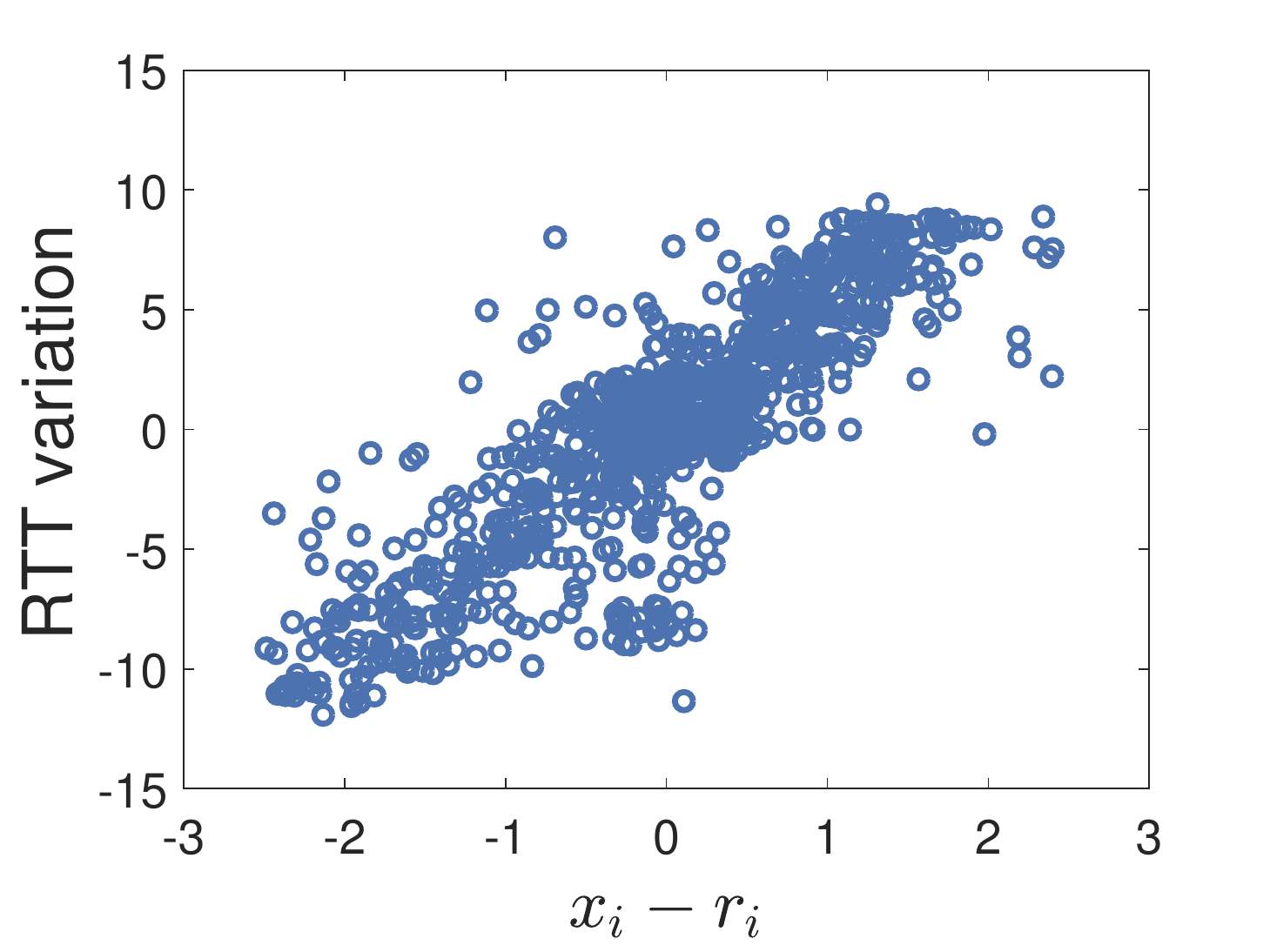}   
\end{minipage}%
}%
\quad             
\subfigure[WiFi network.]{  
\begin{minipage}[t]{0.5\linewidth}  
\centering  
\includegraphics[width=1.7in]{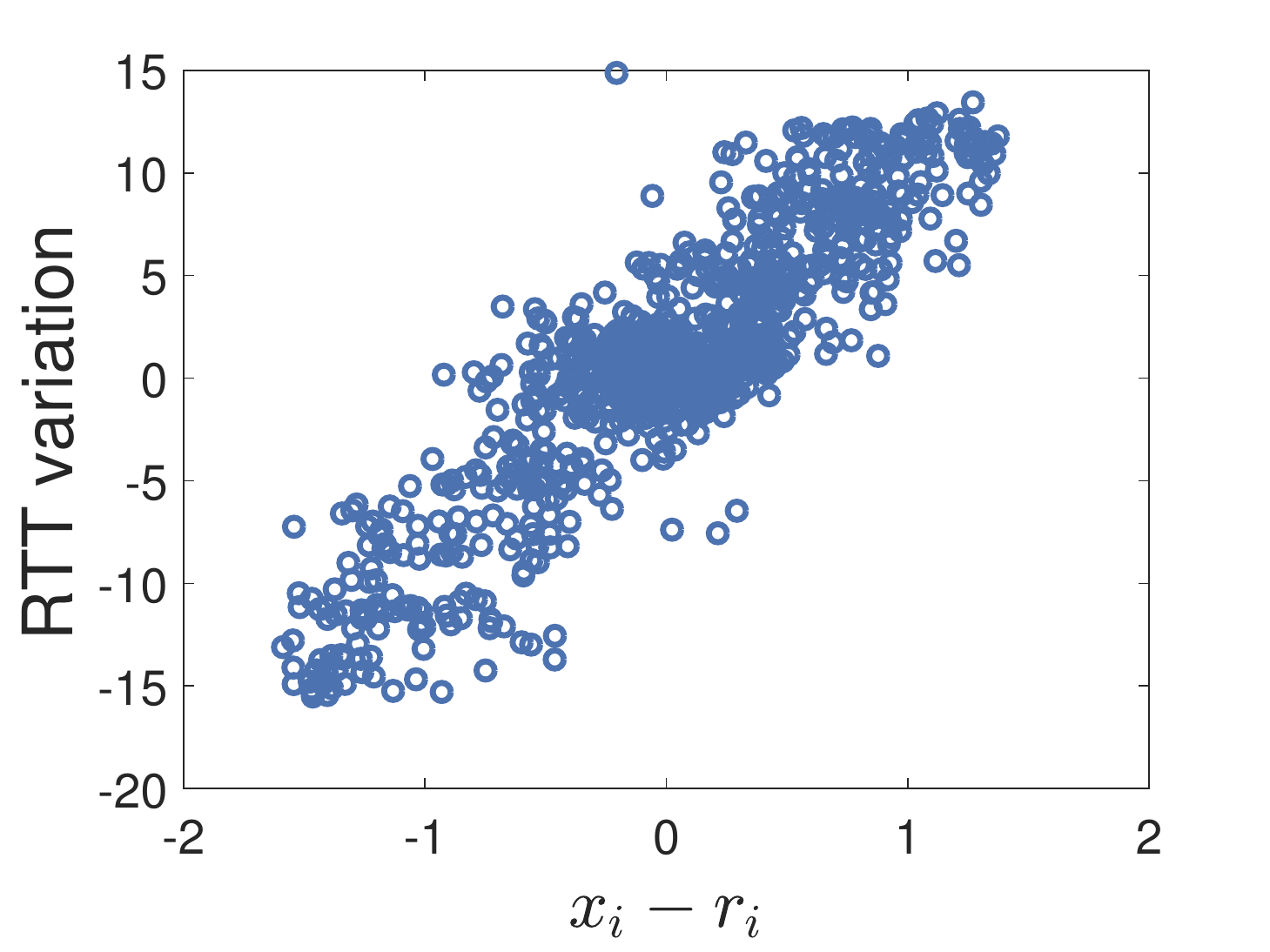}  
\end{minipage}  
}%
\subfigure[LTE network.]{  
\begin{minipage}[t]{0.5\linewidth}  
\centering  
\includegraphics[width=1.7in]{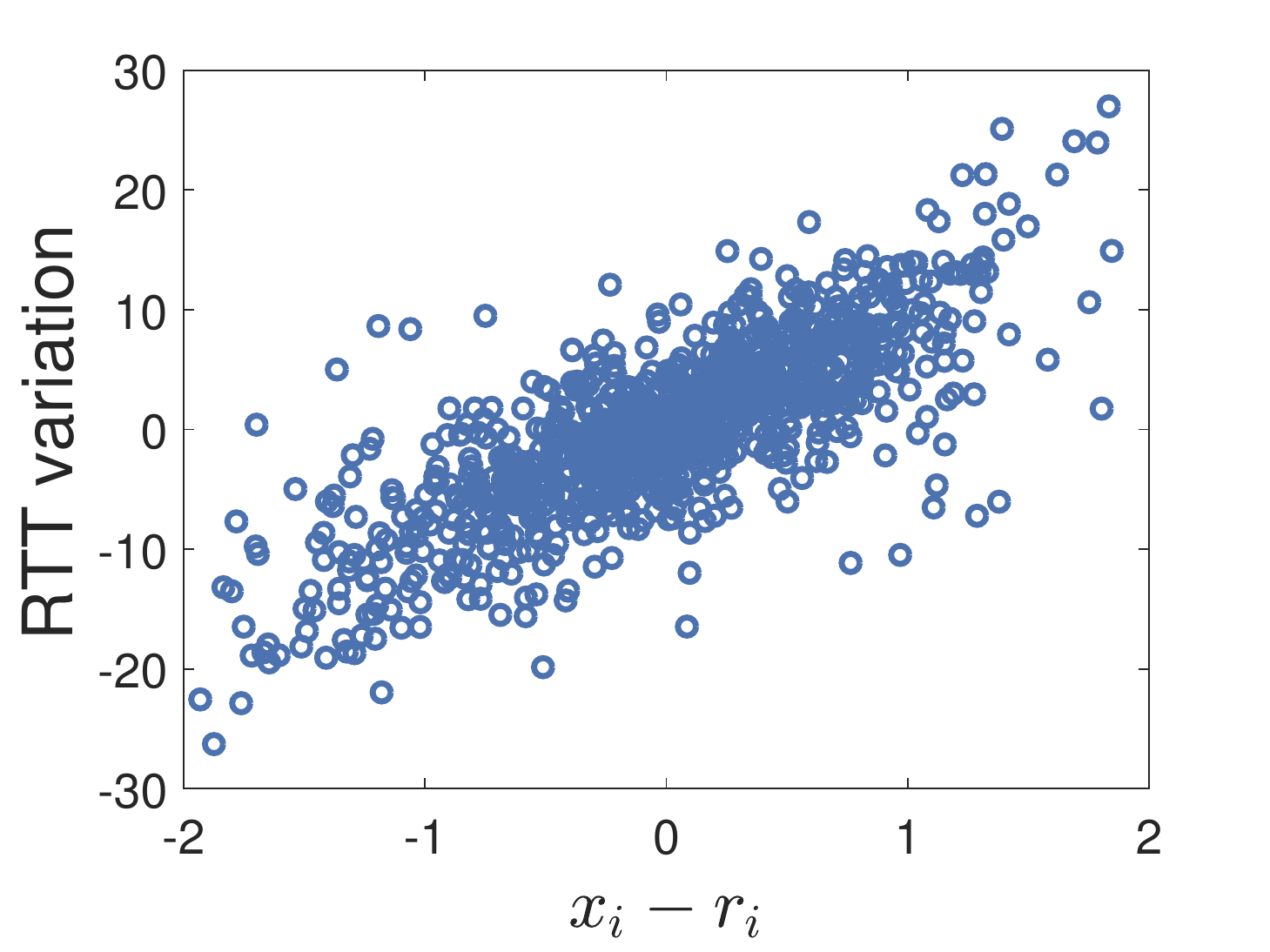}   
\end{minipage}  
}%
\centering  
\caption{The scatter plot of $x - r$ (Mbps) and RTT variation (ms) under different networks.}  
\label{scatter plot}
\end{figure}

In order to understand the dynamics of the network, we generate packets to probe real-world networks, and explore the specific factors that directly affect the network state. To collect the network trace, we build a measurement testbed, consisting of an Ali Elastic Compute Service (ECS), a laptop and a Huawei P20 mobile phone. With UDP used in the transport layer, we tagged packets with sequence numbers and sender timestamp, and implemented ACK return for each packet in the application layer, to enable RTT and receiving rate calculation. The network trace is collected separately under LAN, WAN, WiFi and LTE networks, and the network operators are all China Mobile.

For collecting network trace, UDP packets are emitted with the fixed interval, a sending epoch of 100ms. Within $i$-th epoch, the sending rate $x_{i}$ is constant. The corresponding RTT $rtt_i$ and receiving rate $r_{i}$ is calculated from the ACKs of the all packets emitted in this epoch. The calculation details are shown in \Cref{Receiving}.

\subsection{Correlation Between Sending, Receiving Rate and RTT}
\label{Correlation Analysis}

We collected the trace under different networks, and extracted the sending rate, receiving rate and delay information for correlation analysis. Define $\Delta rtt_{i}$ as the RTT variation between two adjacent epochs:

\begin{equation}
	\Delta rtt_{i} = rtt_{i} - rtt_{i-1}
	\label{deltaRTT}
\end{equation}

\noindent which represents the change of network congestion. 

The correlation analysis is used to study the relationship among sending rate $x$, receiving rate $r$ and RTT variation. \Cref{scatter plot} displays the scatter plots of the traces under different networks, each of which contains more than 1500 data points. Intuitively, whether it is in WAN, LAN, WiFi or LTE networks, the difference between sending rate and receiving rate, i.e. $x-r$, always has a strong positive correlation with $\Delta rtt$. Besides, the correlation is the strongest in LAN, while the correlation decreases slightly in LTE network because of more network jitter noise. This positive correlation indicates that, if the sending rate is larger than the receiving rate, the RTT in the network will also increase, which is not conducive to avoiding congestion.

\subsection{Quantitative Correlation Study}

To verify the correlation, we collected a large number of traces under different networks, respectively in leisure hours (i.e. 9:00 to 11:00) and peak hours (i.e. 19:00 to 21:00). For each trace, Pearson linear correlation coefficient (PLCC) is calculated for statistical analysis.

\begin{figure}[t]
\centering
\includegraphics[width=2.8in]{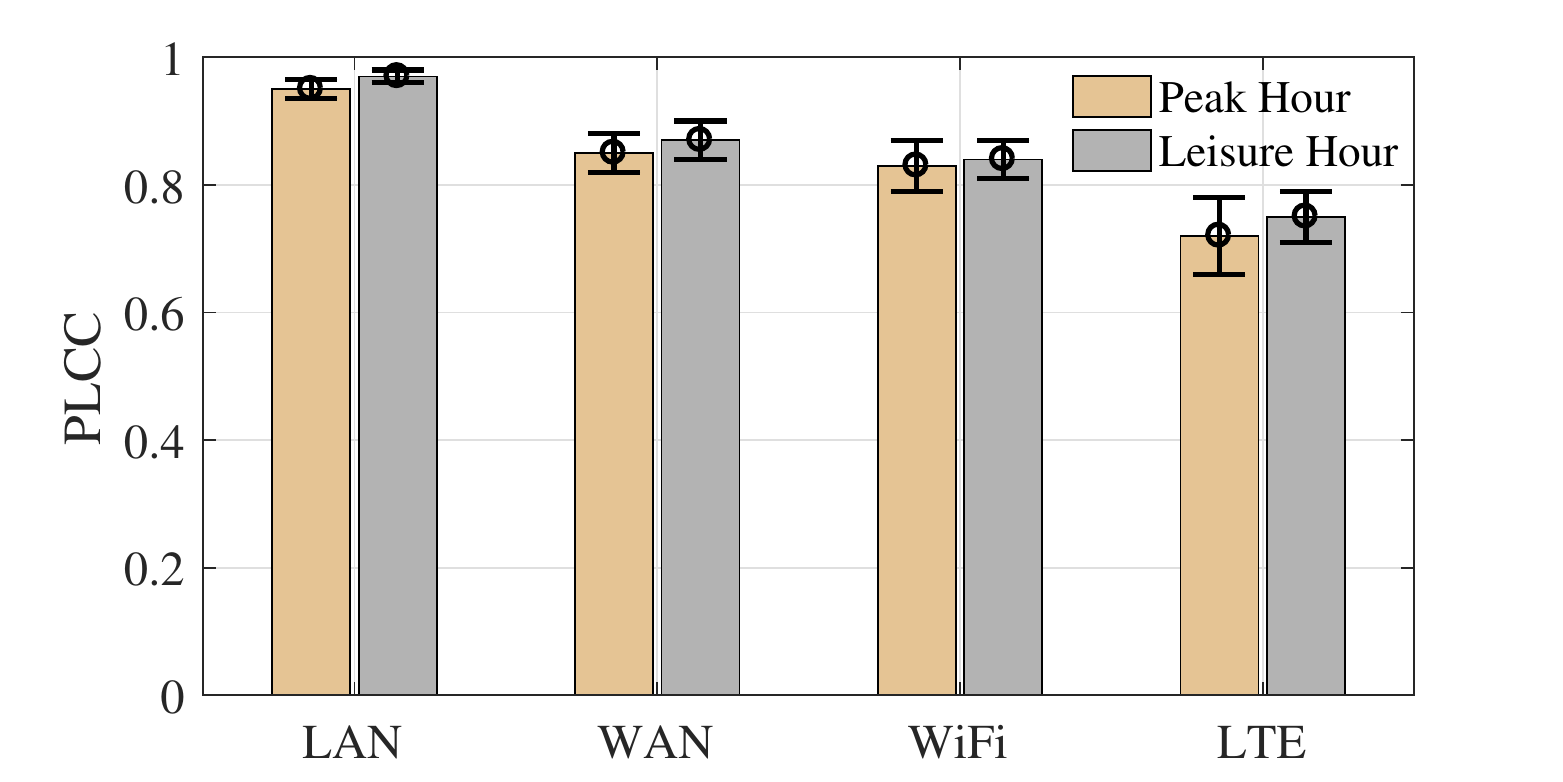}
\caption{Average PLCC of $\Delta rtt$ and $x-r$ with 90\% confidence interval in different real-world networks in leisure hours and peak hours.}
\label{correlation coefficient}
\end{figure}

\Cref{correlation coefficient} displays the PLCC of $\Delta rtt$ and $x-r$ with 90\% confidence interval under different networks in peak and leisure hours. We find that PLCC is the largest in LAN (up to 0.97), but the smallest in LTE networks (around 0.7), which is similar to the discovery in \Cref{Correlation Analysis}. It is mainly because, compared to the simple network environment in the LAN, LTE networks have more noise and will weaken the correlation. Besides, PLCC in leisure hours is a little bigger than that during peak hours, resulting from the fact that the network load is heavier during the peak hour and will produce more noise. 

In summary, PLCC is different under different networks and at different times. But even so, PLCC is still larger than 0.65 in the worst case, which is enough to prove the correlation. Therefore, the further verified conclusion is obtained that \emph{$\Delta rtt$ and $x-r$ have strong linear correlation}. 


\subsection{Linear Regression Learning}
\label{Regression}

Due to the consistent high PLCC as shown in \Cref{correlation coefficient}, we believe the rate difference $x-r$ and RTT variation are always strongly linearly correlated. Therefore, the discovery motivates us to build the model expressing $\Delta rtt_{i}$ as follows:

\begin{equation}
	\Delta rtt_{i} = k \times (x_{i}-r_{i}) +b
	\label{correlation}
\end{equation}

\noindent where $k$ indicates the extent to which $\Delta rtt$ is affected by $x-r$ and $b$ is a bias. Since $b$ can be regarded as a Gauss white noise, we use \emph{maximum likelihood} to solve \Cref{correlation} based on historical data \cite{likelihood-linear}.

This expression is logical from the view of congestion control. When sending rate exceeds receiving rate, the number of packets in the queue of bottleneck link will increase, resulting in the rise of queuing delay and RTT, and vice versa.


\subsection{Distribution of k and b}
\label{Dynamic characteristic}

\begin{figure}[t]  
\centering       
\subfigure[k values in peak hours.]{  
\begin{minipage}[t]{0.5\linewidth}
\centering  
\includegraphics[width=1.7in]{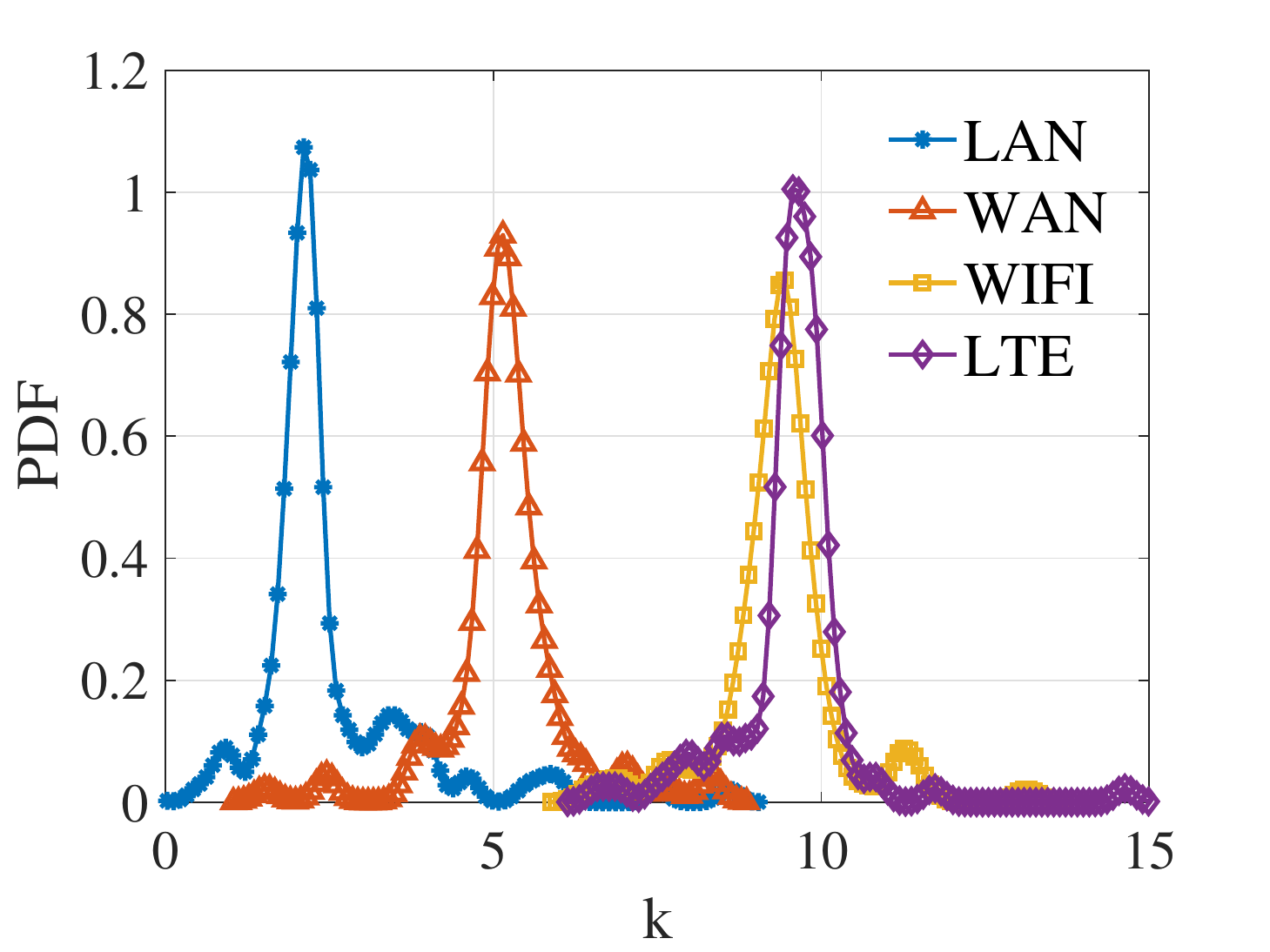}    
\end{minipage}%
\label{cdf_ka}
}%
\subfigure[b values in peak hours.]{  
\begin{minipage}[t]{0.5\linewidth}  
\centering  
\includegraphics[width=1.7in]{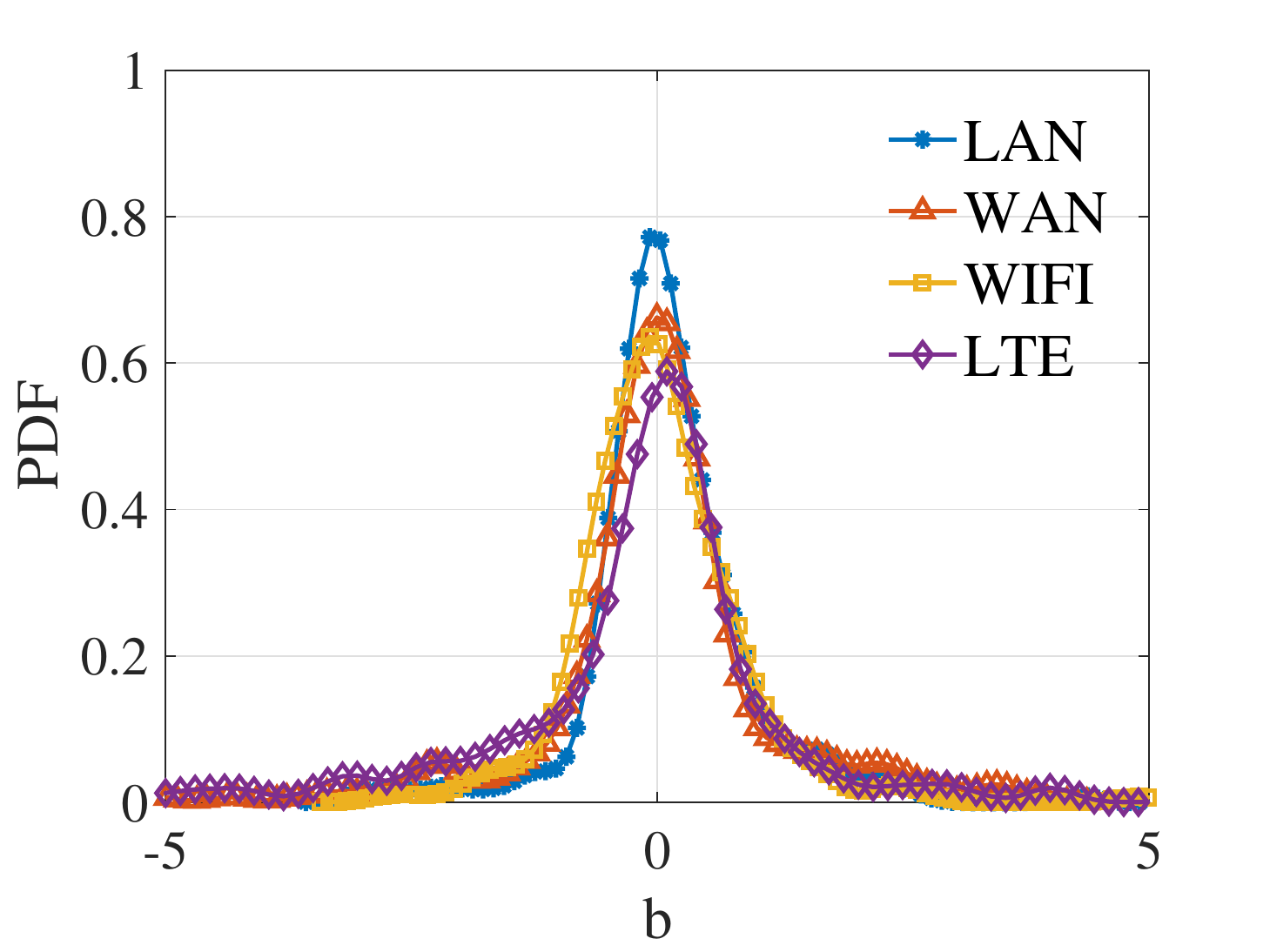}   
\end{minipage}%
\label{cdf_kb}
}%
\quad             
\subfigure[k values in leisure hours.]{  
\begin{minipage}[t]{0.5\linewidth}  
\centering  
\includegraphics[width=1.7in]{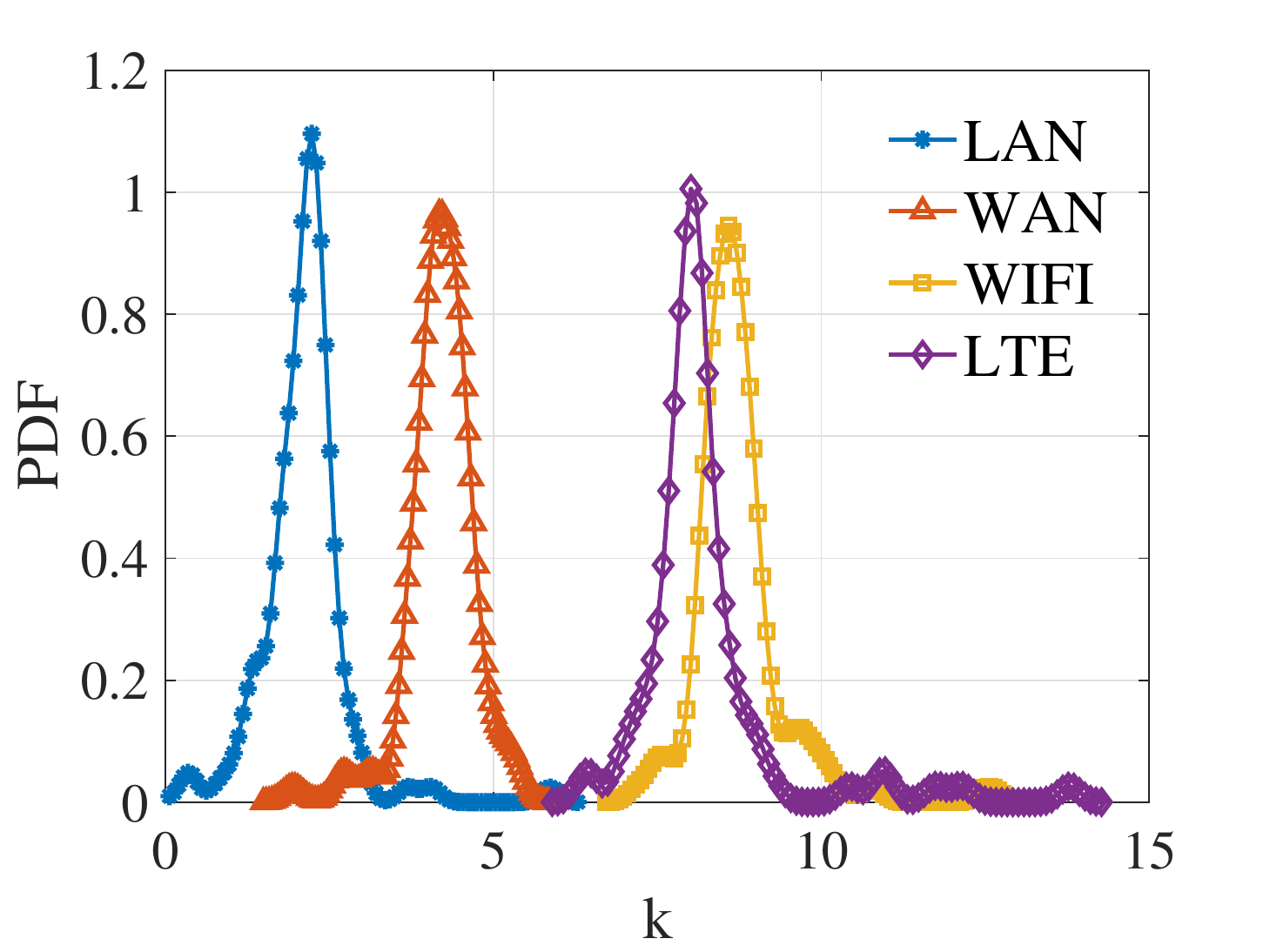}  
\end{minipage}  
\label{cdf_kc}
}%
\subfigure[b values in leisure hours.]{  
\begin{minipage}[t]{0.5\linewidth}  
\centering  
\includegraphics[width=1.7in]{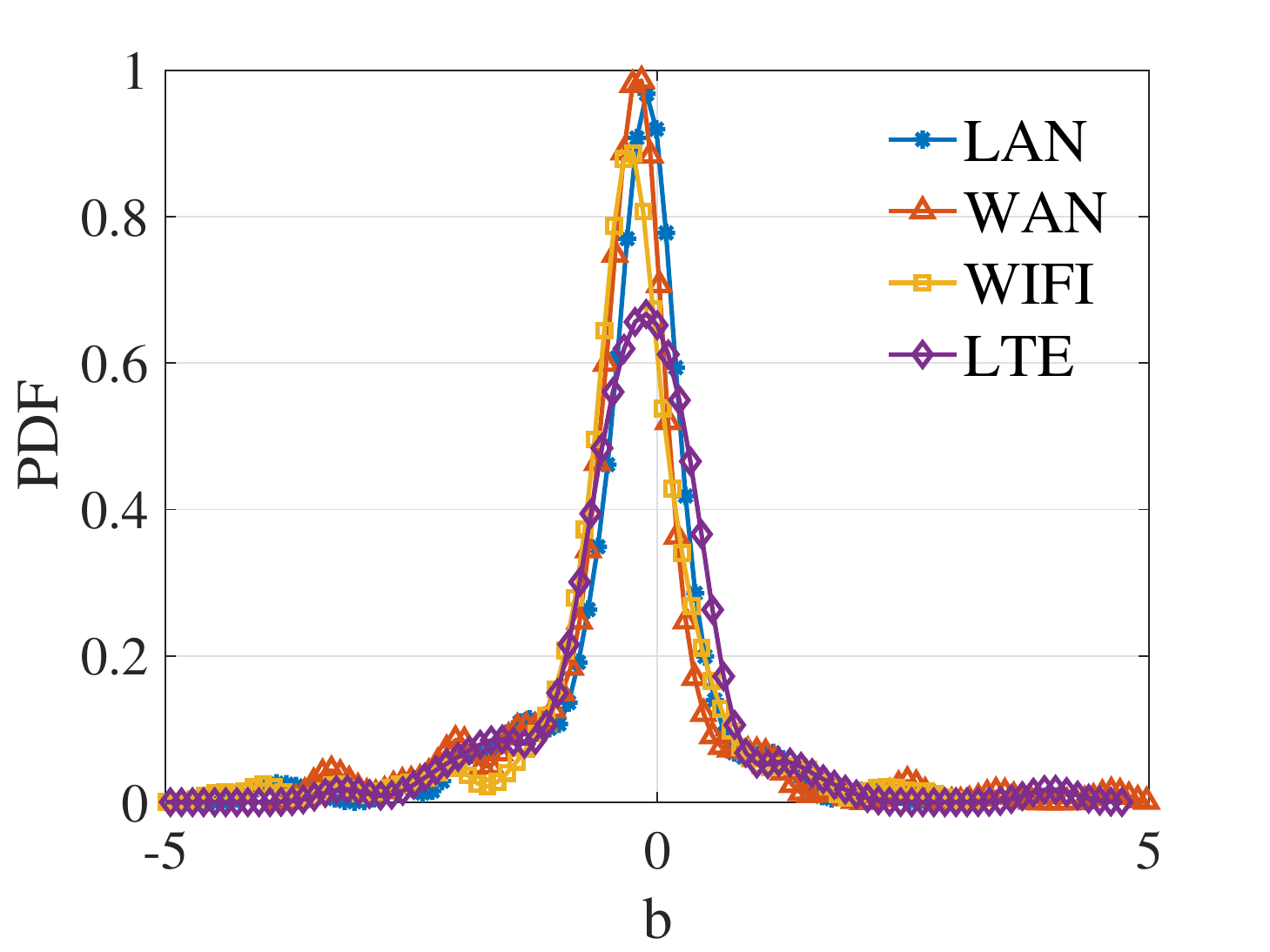}   
\end{minipage}  
\label{cdf_kd}
}%
\centering  
\caption{The probability density function (PDF) of $k$ and $b$ values under different networks in leisure and peak hours.}  
\label{cdf_k}
\end{figure}

Furthermore, we study whether this function expression is constant in the real complex networks. Based on the collected traces, above maximum likelihood estimation is used to obtain the quantized values of $k$ and $b$. \Cref{cdf_k} displays the probability density function (PDF) of obtained $k$ and $b$ under different networks, respectively in leisure hours (i.e. 9:00 to 11:00) and peak hours (i.e. 19:00 to 21:00). It shows that the distributions of $k$ and $b$ are always approximate Gauss distribution regardless of the network type, which validates the applicability of the maximum likelihood method.

For further analysis, under different networks, the mean of $k$ is obviously different, which is mainly related to the available bandwidth at that time. But as for $b$, the mean value is always near 0, no matter what type of network it is in. Although its variance is slightly different, we think of $b$ as a Gauss noise, and it is usually small enough to be overlooked.



As shown in \Cref{cdf_ka} and \Cref{cdf_kc}, although the distribution of $k$ values is always Gaussian, its mean value has shifted markedly. That is to say, the distribution of $k$ values will change over time in the same network. Therefore, the way to generate a fixed function to adjust the sending rate is not optimal. Because when the network conditions change, the functional relationship between $x-r$ and $\Delta rtt$ will also change. It inspires us to design the step size adaptation mechanism as described in \Cref{k value update}.  

\subsection{Key Insights}

Through the above correlation analysis and verification, we further emphasize the motivation of this paper into three aspects:

\begin{itemize} 

  \item An important discovery is observed, i.e. the difference between sending rate $x$ and receiving rate $r$ always has a strong correlation with RTT variation $\Delta rtt$.
  
  \item Through the analysis of correlation coefficient, the linear model as \Cref{correlation} is established, which enlightens us on congestion control model.
  
  \item The values of $k$ and $b$ that represents the model parameters are various at different networks and different times, so a step size adaptation mechanism is necessary.
  
\end{itemize}

\section{Learning Based Congestion Control}
\label{algorithm}

In this section, our Iris is introduced, which is a statistical learning based congestion control deeply inspired by our research on the correlation analysis in real-world networks. 

\subsection{Iris Algorithm}

\subsubsection{Key Ideas} \
\label{goal}

Iris is designed as an efficient real-time congestion control, which mainly consists of two modules, a \textbf{low-latency fairness model} and a \textbf{learning-based rate adjustment}. The fairness model forces all flows to keep a small and constant number of packets queuing in network, i.e. \emph{queue load}, so as to achieve fairness and low latency concurrently when there are multiple flows competing on a bottleneck link. The learning-based rate adjustment builds a statistical function between RTT and rate through online linear regression learning, then uses the model to determine the proper sending rate. It avoids fixed adjustment step size and converges to the fairness objective more quickly.  \\

\subsubsection{Low-latency fairness Model} \
\label{Model}

As the core of Iris, the objective function with the ability to achieve low latency and fairness is designed, and its main component is the estimation of \emph{queue load}. 


As mentioned above, $rtt_{i}$ represents the average RTT within the $i$-th epoch. We define $T$ as target delay, which is set to the minimum RTT in a time window by default.  So $rtt_{i}-T$ represents the extra queuing delay and the number of extra packets in bottleneck queue within $i$-th epoch is calculated as $x_{i} \cdot (rtt_{i}-T)$. It represents the extra queue load estimation which indicates the extent of congestion. Therefore, the congestion can be avoided by maintaining it at a  fixed range, i.e. $x_{i} \cdot (rtt_{i}-T) = B$, with $B$ defined as the queue load target.



Therefore, our objective function can be expressed as:

\begin{equation}
	U(x_{i}, rtt_{i})=x_{i} \cdot (rtt_{i}-T) - B
	\label{convergence point}
\end{equation}

\noindent where $B$ represents the convergence status of Iris. Fewer redundant packets in the bottleneck queue helps Iris achieve high channel utilization and low latency at the same time. Thus it also tells the sender how to adjust its sending rate. When $x_{i} \cdot (rtt_{i}-T)$ is higher than $B$, the network is considered to be congested, so the sending rate is supposed to be decreased,  and vice versa. In addition, the fairness is guaranteed if each flow maintains the same number of packets filled into the bottleneck link.

\subsubsection{Expected RTT Variation Adjustment} \
\label{Rate Estimator}


The objective function in our \emph{Low-latency Fairness Model} implies the extent of congestion. Combining the results of our previous correlation analysis, we designed the following strategies to adjust the expected next RTT Variation $d_{i+1}$: when $U(x_{i}, rtt_{i}) < 0$, the queue load does not reach our target. To improve bandwidth utilization, next RTT Variation $d_{i+1}$ is expected to be greater than 0,  because it usually indicates a higher sending rate. In the opposite case, $d_{i+1}$ is expected to be less than 0. By adjusting the expected RTT step by step, \Cref{convergence point} will tend to be established.  Therefore, we first simply define the expected RTT variation for the next epoch $d_{i+1}$ as follows:

\begin{equation}
	d_{i+1}=-U(x_{i}, rtt_{i})
	\label{delta_d}
\end{equation}





However, the real-world network has jitter noise so that the measurement of RTT is not completely accurate, which also leads to the existence of abnormal value of $U(x_{i}, rtt_{i})$. To prevent over-adjustment, the activation function $tanh$ is introduced to limit the range of RTT variation, which is expressed as follows.

\begin{equation}
	tanh(U) = \frac{e^{U}-e^{-U}}{e^{U}+e^{-U}}
	\label{tanh}
\end{equation}

\noindent The main reason for using $tanh$ function is that it has an effective suppression on abnormal values. However, in the original $tanh$ function, the effective range of independent variables is only about [-1,1], while our objective function U can be as high as dozens. Therefore, in the process of implementation, we also need to stretch the effective scope of $tanh$ function through dividing $U(x_{i}, rtt_{i})$ by an expansion coefficient $M$. Finally, $d_{i+1}$ can be expressed as follows.

\begin{equation}
	d_{i+1}=-\delta \cdot tanh(\frac{U(x_{i}, rtt_{i})}{M})
	\label{delta_d}
\end{equation}

\noindent where $\delta > 0$ represents the threshold of RTT variation.

By this method, the farther away from convergence target, the larger adjustment step size will be adopted. Besides, it helps to solve the problem of delay overestimation \cite{vegas-overestimate}.\\

\subsubsection{Learning-based Rate Adjustment} \
\label{Rate Adjustment}

According to the result of correlation analysis in \Cref{motivation}, the strong linear correlation between $\Delta rtt$ and $x-r$ is captured. Therefore, if the expected next RTT Variation $d_{i+1}$ has been determined, the next sending rate $x_{i+1}$ can be expressed as:

\begin{equation}
	x_{i+1} = r_{i+1} + \frac{d_{i+1}}{k}
	\label{final}
\end{equation}

\noindent  but when Iris makes the decision on $x_{i+1}$ for the next epoch, $r_{i+1}$ is unknown to the sender, so it is estimated by recently seen receiving rate in implementation, i.e. $r_{i}$, and $k$ is learned from linear regression of historical data, determining the rate adjustment step size. 

The design of \Cref{final} is reasonable. When $d_{i+1}$ is greater than 0, the channel is considered underutilized and we tend to fill more packets into the bottleneck queue, so the sending rate is supposed to be increased. On the contrary, if $d_{i+1}$ is less than 0, it means that the number of packets in the queue exceeds the budget and is easy to cause congestion. In this case, sending rate is supposed to be adjusted to the decreasing direction.

\subsection{Fairness Analysis}

Iris uses the low-latency fairness model, coupling with the learning-based rate adjustment, to guarantee high performance from the individual sender's perspective and ensure the convergence to a global fair rate allocation. Specifically, we consider a network model in which $N$ flows compete on a bottleneck link with the bandwidth of $C$ and the buffer is a FIFO queue.

\textbf{Theorem.} \emph{When N Iris-senders share a congested bottleneck link and each sender uses the rate control mechanism as \Cref{final}, the senders' sending rates converge to a global fair configuration.}

\noindent \emph{Proof.} We define that for a pair of senders $a$ and $b$, within epoch $i$, their sending rates are respectively $x_{i,a}$ and $x_{i,b}$. And $x_{i,b}$ is larger than $x_{i,a}$. So, if the rate difference between any two flows decreases with time, the theorem is valid. It is equivalent to proving that

\begin{align}
 |x_{i+1,b}-x_{i+1,a}| &< |x_{i,b}-x_{i,a}| \notag \\
 |r_{i,b}-r_{i,a} +\frac{d_{i+1,b}}{k} -\frac{d_{i+1,a}}{k}| &<|x_{i,b}-x_{i,a}| 
 \label{leq}
\end{align}

The bottleneck link is congested, i.e. $\sum_{j \in N} x_{i,j}>C$. Combined with the network link model, the difference in receiving rate can be expressed as $r_{i,b}-r_{i,a}=\frac{C (x_{i,b}-x_{i,a})}{\sum_{j \in N} x_{i,j}}$. Therefore, the following formula can be obtained.

\begin{equation}
	 0<r_{i,b}-r_{i,a} <x_{i,b}-x_{i,a}
\end{equation}

In order for the condition of \Cref{leq} to be satisfied, the following equation must be established.

\begin{align}
	 0<\frac{d_{i+1,a}}{k} -\frac{d_{i+1,b}}{k} <x_{i,b}-x_{i,a}  
\end{align}

As the buffer is shared across all flows on the bottleneck link, they have the same queuing delay, i.e. $q_{i}=rtt_{i,a}-T_{i,a}=rtt_{i,b}-T_{i,b}$. And the $tanh$ function in \Cref{delta_d} is monotonically increasing, hence $\frac{d_{i+1,a}}{k}>\frac{d_{i+1,b}}{k}$. Defining $U_{i}$ as  $\frac{x_{i}(rtt_{i}-T)-B}{M}$, we can prove that


\begin{align}
 \frac{1}{k}(d_{i+1,a} - d_{i+1,b}) &< x_{i,b}-x_{i,a} \notag \\
 \frac{\delta}{k} (tanh(U_{i,b})-tanh(U_{i,a})) &<x_{i,b}-x_{i,a} \notag \\
 \frac{\delta}{k} \frac{tanh(U_{i,b})-tanh(U_{i,a})}{U_{i,b}-U_{i,a}} \frac{U_{i,b}-U_{i,a}}{x_{i,b}-x_{i,a}} &<1 \notag \\
 \frac{\delta}{k} \frac{rtt_{i}-T}{M} \frac{tanh(U_{i,b})-tanh(U_{i,a})}{U_{i,b}-U_{i,a}} &<1
 \label{gradient}
\end{align}

\noindent where  $tanh$ function has the largest slope at zero and its value is 1, so $0<\frac{tanh(U_{i,b})-tanh(U_{i,a})}{U_{i,b}-U_{i,a}}<1$. And the value of $M*k$ is usually much larger than $\delta(rtt_{i}-T)$, as described in \Cref{pp}. Thus the \Cref{leq} is established and the fair configuration is achieved.

%


\section{Implement}
\label{Implement}

\subsection{Receiving Rate Estimation}
\label{Receiving}

\begin{figure}[t]
\centering
\includegraphics[width=3in]{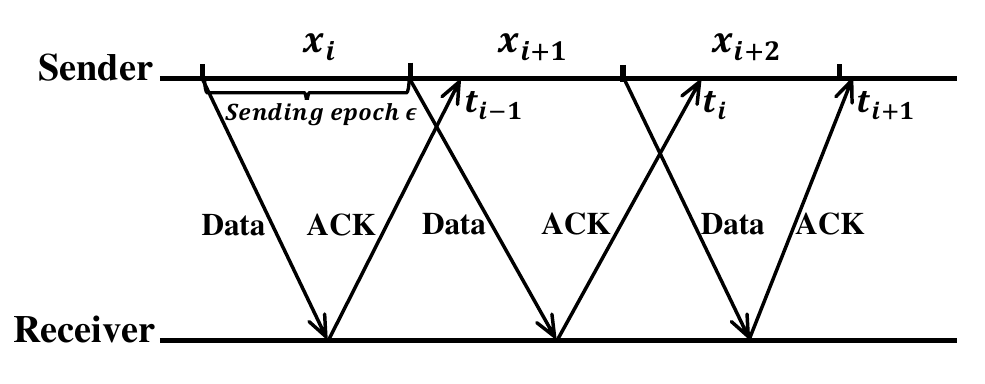}
\caption{A diagram for calculating the receiving rate.}
\label{receive_node}
\end{figure}

Considering that the channel unpredictably changes over time, especially in wireless networks, Iris adjusts sending rate in a small epoch of $\epsilon$ ms to quickly adapt to the changing link. In order to obtain the receiving rate estimation, we first tagged packets with sequence numbers and sender's timestamp, and implemented ACK return for each data packet in the application layer, with UDP used in the transport layer.

In particular, our receiving rate is not obtained by the common calculation method. As shown in \Cref{receive_node}, within $i$-th epoch, the total number of packets sent is $x_{i} \cdot \epsilon$. $t_{i}$ represents the ACK return time of the last packet of this epoch, thus $t_{i}-t_{i-1}$ represents the total time for the successful delivery of all packets in the $i$-th epoch. So the receiving rate estimate $r_{i}$ of the $i$-th epoch is calculated at $t_{i}$, as follows.

\begin{equation}
	 r_{i}= \frac{x_{i} \cdot \epsilon}{t_{i}-t_{i-1}}
\end{equation}

\subsection{Update k}
\label{k value update}

As described in Section \ref{Dynamic characteristic}, the $k$ value reflects the relationship between $\Delta rtt$ and $x-r$, which directly determines the rate adjustment step size according to \Cref{final}. But it will change with the dynamics of the network. That is to say, a fixed k value can not be effective under various network conditions. Therefore, a periodic update mechanism for the $k$ values is designed in Iris. To keep up with the network dynamics, the update cycle is set to 5 seconds empirically. Then, linear regression learning is used to fit $k$ values periodically, based on the historical data.


\subsection{Cold Start}

At the start-up of Iris, there is no data to support it to obtain the effective $k$ value, and any handcrafted initial value is difficult to be robust to all kinds of networks with different capacities. Therefore, we adopt the similar control strategy as the slow-start stage of TCP to quickly perceive the link capacity and collect the training data in this cold start process. 

After Iris starts, the initial sending rate of 100Kbps is first adopted, then it is doubled every epoch for updating the next sending rate. Once the packet loss rate is increased, Iris will exit from the startup phase and use the collected data to learn an initial $k$ value. After that, $k$ is periodically updated as described in \Cref{k value update}.

\subsection{Parameters Settings}
\label{pp}

\begin{table}[t]
\caption{Default parameter settings.}
\centering
\scalebox{1}{
\begin{tabular}{cc}
\toprule[1.1pt]
Parameter& Value\\
\midrule
Epoch time $\epsilon$ ms & 50 \\
Scaling multiplier $M$ of $tanh$ & 100\\
Extra queue load target $B$ & 10\\
RTT variance boundary $\delta$ & 3\\
\bottomrule[1.1pt]
\end{tabular}}
\label{Parameters}
\end{table}

Different parameters will directly affect the performance of iris. Unless stated otherwise, we implement Iris using the parameter default values in \Cref{Parameters}. $\epsilon$ represents the sending rate adjustment interval. The smaller $\epsilon$ is more conducive to improving the adaptability of the algorithm, but will affect the accuracy of receiving rate estimation in a single epoch. Considering these two aspects, we empirically set $\epsilon$ as 50ms based on a large number of experimental tests under different networks. $M$ determines the the effective limits of $tanh$ function and $\delta$ limits the range of RTT variance, empirically set to 100 and 3 respectively, taking into account the applicability and generality for various networks. $B$ is the extra queue load target and too large $B$ will result in high self-inflicted delay, so we set the value of $B$ to 10.

\section{Performance Evaluation}
\label{Evaluation}

In this section, we conducted extensive experiments to evaluate the performance of Iris, considering both the transport layer and the application layer. The experimental environment consists of a variety of networks, including emulated network, real-world Internet and commercial LTE networks.

\subsection{Testbed}
\label{testbed}

In order to evaluate the performance of Iris in transport layer and network layer respectively, we implement it in UDT \cite{udt, udtcode} and QUIC \cite{quic}. The aim of QUIC implementation is to build an HTTP live streaming server, to evaluate the bitrate and PSNR gain of Iris in the application layer.

For transport layer testing, we implement a user-space prototype based on UDT. In this scenario, the comparison objects mainly include Sprout \cite{sprout,sproutcode}, PCC (with latency utility function by default) \cite{pcccode}, BBR and TCP variants (e.g. Cubic and Vegas). Two main metrics of the performance are considered, namely throughput and delay characteristics. The testbeds in emulation and real-world environment are introduced below.

In the emulation environment, we design a dumbbell topology consisting of two nodes in the LAN, where one plays as sender and the other is receiver. The nodes are connected through Gigabit fiber and run a configurable number of sources with \emph{Linux core 4.9}. For PCC and Sprout, the developers' implementations are used. We employ the \emph{NetEm} linux module along with the traffic shaper \emph{TC} to configure the link parameters, such as bottleneck bandwidth, propagation delay, packet loss and maximum queue size. Besides, \emph{tcpdump} is used to capture packets to measure the metrics. 

In the real-world environment, we mainly consider intercontinental Ethernet and LTE mobile network. In the case of intercontinental Ethernet, four Ali ECSs are employed to establish a node in Beijing, HongKong, Singapore and America respectively. Our host is used as sender while these nodes are regarded as receivers. As for LTE network, we employ a laptop connected with a 4G mobile phone to run these algorithms on China Mobile, a commercial LTE network. In order to evaluate the performance of these algorithms, a lot of tests have been carried out considering different time, different network access modes and other factors. The total time of collected traces is over 50 hours.

For the evaluation in application layer, in addition to the HTTP live streaming server over QUIC, we implement a  a video player based on the open source MPEG-DASH \emph{dash.js} \cite{dashjs}. The QUIC server provides the video with five different bitrate versions (350Kbps, 620Kbps, 1.57Mbps, 2.54Mbps and 3.60Mbps), which are divided into segments with the same length (1 second), and the buffer length is set as 15 seconds. The traffic shaper \emph{TC} is also used to configure the link parameters from server to client. When switching different algorithms in the congestion control module of QUIC, the performance of Iris can be evaluated based on the information output from the browser console. 

\subsection{Evaluation in Emulated Networks}

\subsubsection{Robustness to Random Loss} \

\begin{figure}[t]
\centering
\includegraphics[width=2.8in]{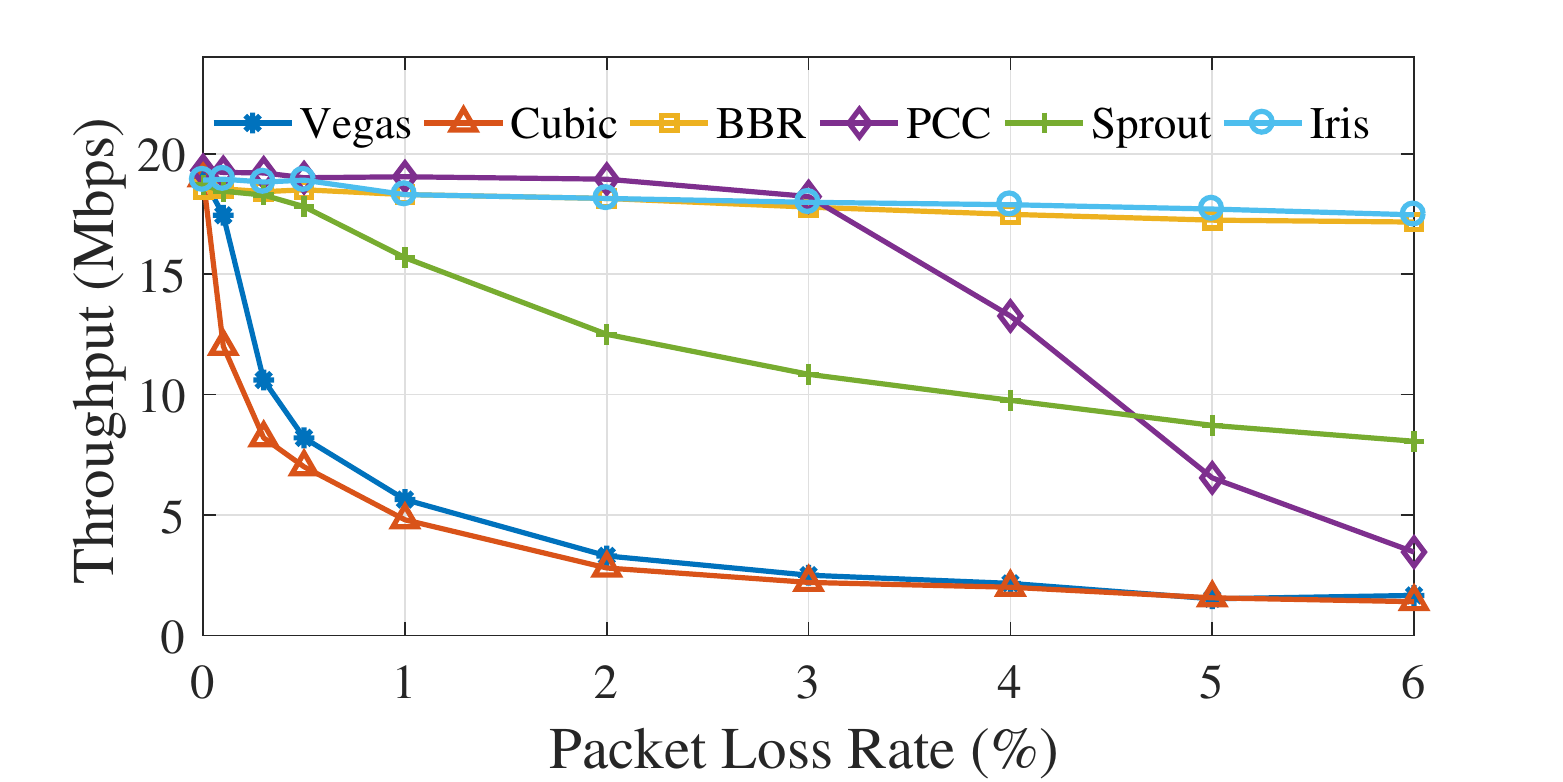}
\caption{Iris is robust to random packet loss.}
\label{fig:loss-compare}
\end{figure}

Lossy links in today’s networks are not uncommon: wireless links are often unreliable and very long wired network paths can also have random loss caused by unreliable infrastructure. To further quantify the effect of random loss, we use \emph{tc} to configure a bottleneck link with 20Mbps bandwidth, 50ms round-trip propagation delay and varying loss rate, to evaluate the algorithms' robustness to random loss.

As Figure \ref{fig:loss-compare} shown, Cubic and Vegas perform badly when there is random packet loss. Even 0.5\% of the random packet loss rate will reduce their channel utilization to less than 50\%. And the performance of PCC will also drop sharply when the packet loss rate is greater than 3\%. In contrast, our Iris and BBR are robust to random packet loss, which is because the packet loss is not considered as a congestion signal in the rate adjustment mechanism. \\

\subsubsection{Tolerance to Long RTT} \

Satellite Internet is widely used for critical missions such as emergency and military communication, which is challenging for congestion control because it has high propagation delay (RTprop), large bandwidth-delay product and random loss. Referring to the real-world measurement for the WINDs satellite Internet system \cite{Satellite}, we evaluate Iris on an emulated link with 20 Mbps capacity, 1 BDP of buffer, 0.74\% stochastic loss rate and changing RTprop. 


\Cref{LongRTT} shows the average throughput and queuing delay of the protocols v.s. Iris. Compared against Iris, the throughput of Cubic and Vegas is less than 10\%, due to the random packet loss in the link. PCC obtained the highest throughput, but at the cost of excessive latency. BBR is able to ignore random loss, but still underutilized the link as its rate oscillated wildly. In contrast, Iris achieves over 70\% of optimal throughput at a time delay cost of only around 20 milliseconds. Although it is also affected by the high BDP, a good trade-off is achieved between throughput and delay. 

\begin{table}[ht]
\caption{Average throughput (Mbps) and queuing delay (ms) vs. Iris in emulated satellite link with high RTprop.}
\centering
\scalebox{1}{
\begin{tabular}{ccccccc}
\toprule[1.1pt]
RTprop &\multicolumn{2}{c}{600ms}&\multicolumn{2}{c}{800ms}&\multicolumn{2}{c}{1000ms}\\
\midrule
Country&Rate&Delay&Rate&Delay&Rate&Delay\\
Iris& 16.7 & 16 & 15.6 & 19 & 13.9 & 21 \\
Vegas &$0.03\times$&$0.06\times$&$0.02\times$&$0.05\times$&$0.03\times$&$0.09\times$\\
Cubic &$0.06\times$&$0.12\times$&$0.03\times$&$0.15\times$&$0.03\times$&$0.19\times$\\
BBR &$0.71\times$&$1.02\times$&$0.65\times$&$0.88\times$&$0.51\times$&$0.93\times$\\
PCC &$1.12\times$ &$31.3\times$&$1.24\times$&$38.6\times$&$1.34\times$&$42.3\times$\\
\bottomrule[1.1pt]
\end{tabular}}
\label{LongRTT}
\end{table}

\subsubsection{Responsiveness to Network Variation} \

We next demonstrate how quickly Iris can adapt to dynamically changing network conditions. We start with a network employing \emph{tc} where the bottleneck bandwidth changes every 40 seconds, with 50ms round-trip propagation delay and 10KB buffer. For each protocol, we repeat the test with 160 seconds duration.

\begin{figure}[t]
\centering
\includegraphics[width=2.8in]{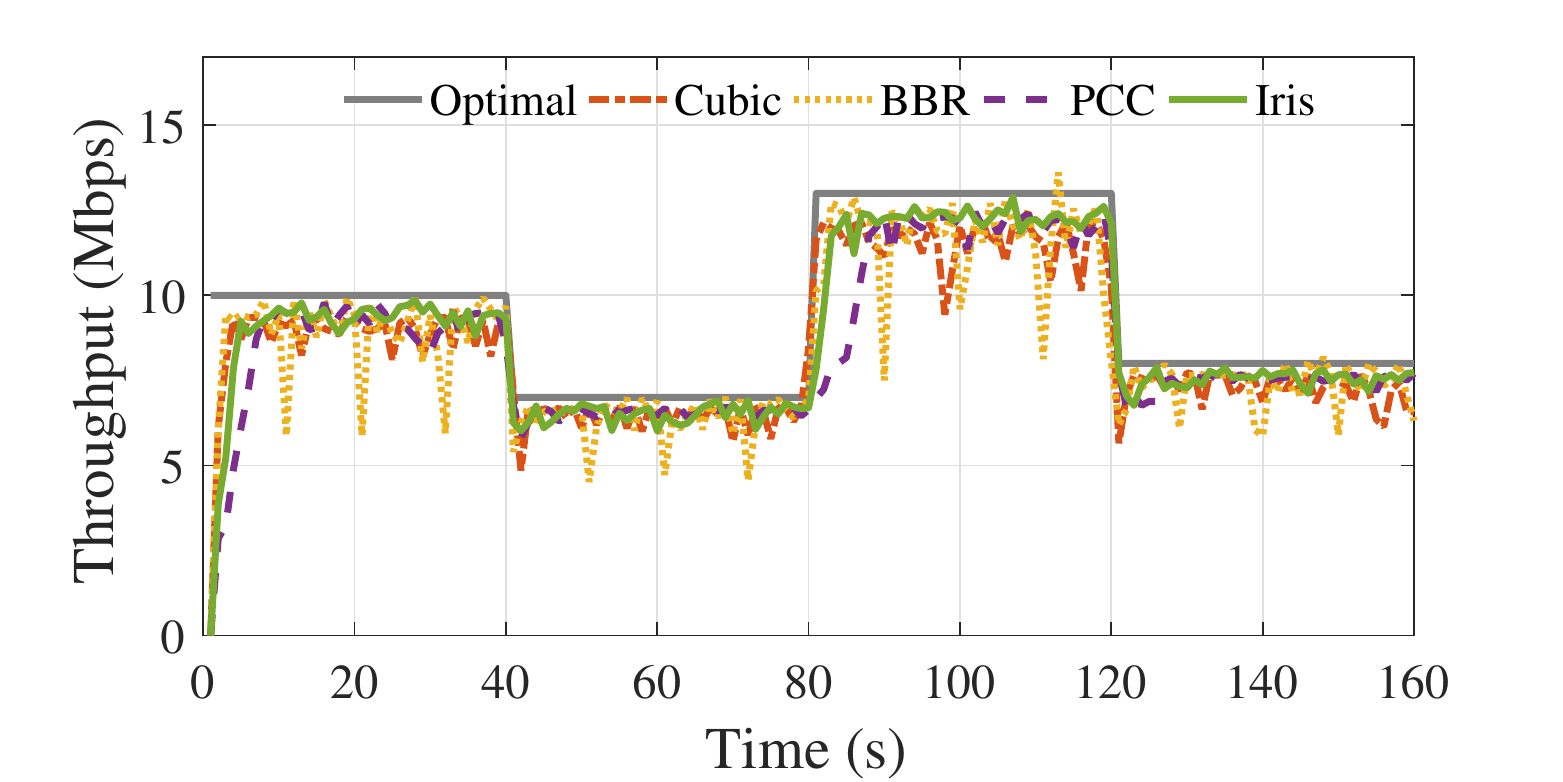}
\caption{Reaction to Changing Network.}
\label{fig:fangbo}
\end{figure}

\Cref{fig:fangbo} illustrates the behavior of several of the protocols across time. Our Iris and PCC have almost the same throughput in the stable state, but when the bottleneck bandwidth changes, PCC's response is much slower. From startup to rate stability, PCC even takes ten seconds. This is mainly because PCC determines the direction of adjustment by several attempts before the sending rate is adjusted. As for Cubic, the higher capacity results in the larger rate jitter, due to the small buffer. BBR will periodically reduce its sending rate because of its time delay synchronization mechanism. \\

\subsubsection{Fairness Evaluation} \
\label{sec:fairness evaluation}

\emph{Intra-protocol fairness} \ 
\label{Intra-protocol}

We first evaluate the intra-protocol fairness of Cubic, BBR, PCC and our Iris separately. For this purpose, a dumbbell topology in the LAN is built to demonstrate their dynamic behavior with three flows sharing a bottleneck link with 20 Mbps bandwidth, 50 ms RTT and 0.5 BDP buffer. 

\begin{figure}[t]  
\centering       
\subfigure[Cubic Self-fairness.]{  
\begin{minipage}[t]{0.5\linewidth}
\centering  
\includegraphics[width=1.6in]{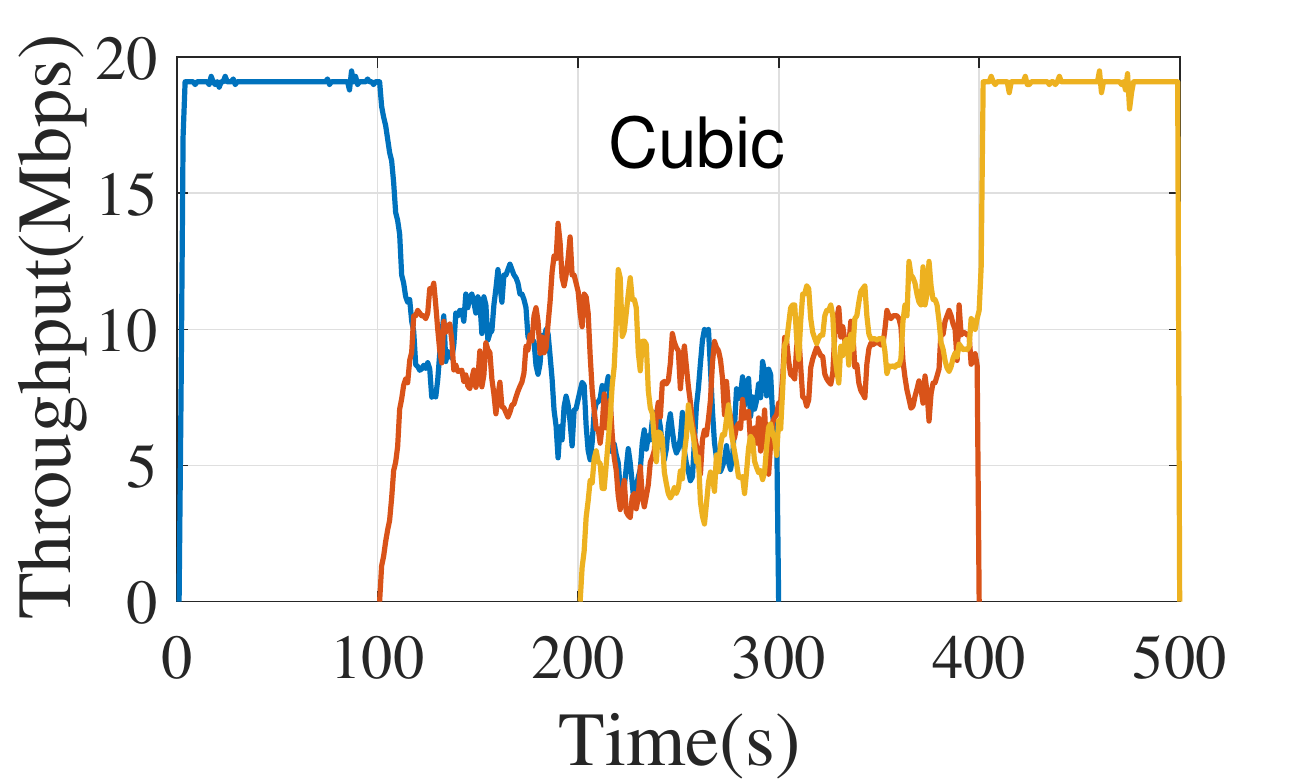}    
\end{minipage}%
}%
\subfigure[BBR Self-fairness.]{  
\begin{minipage}[t]{0.5\linewidth}  
\centering  
\includegraphics[width=1.6in]{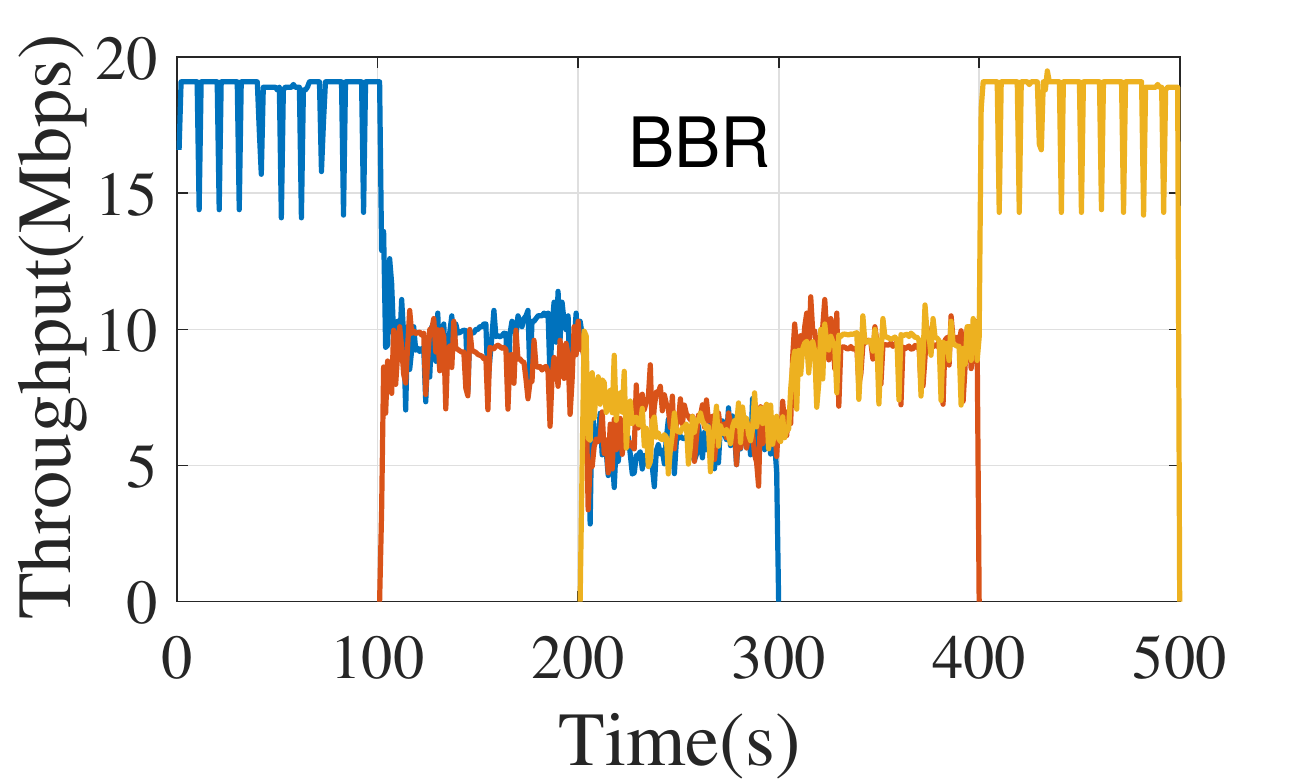}   
\end{minipage}%
}%
\quad            
\subfigure[PCC Self-fairness.]{  
\begin{minipage}[t]{0.5\linewidth}  
\centering  
\includegraphics[width=1.6in]{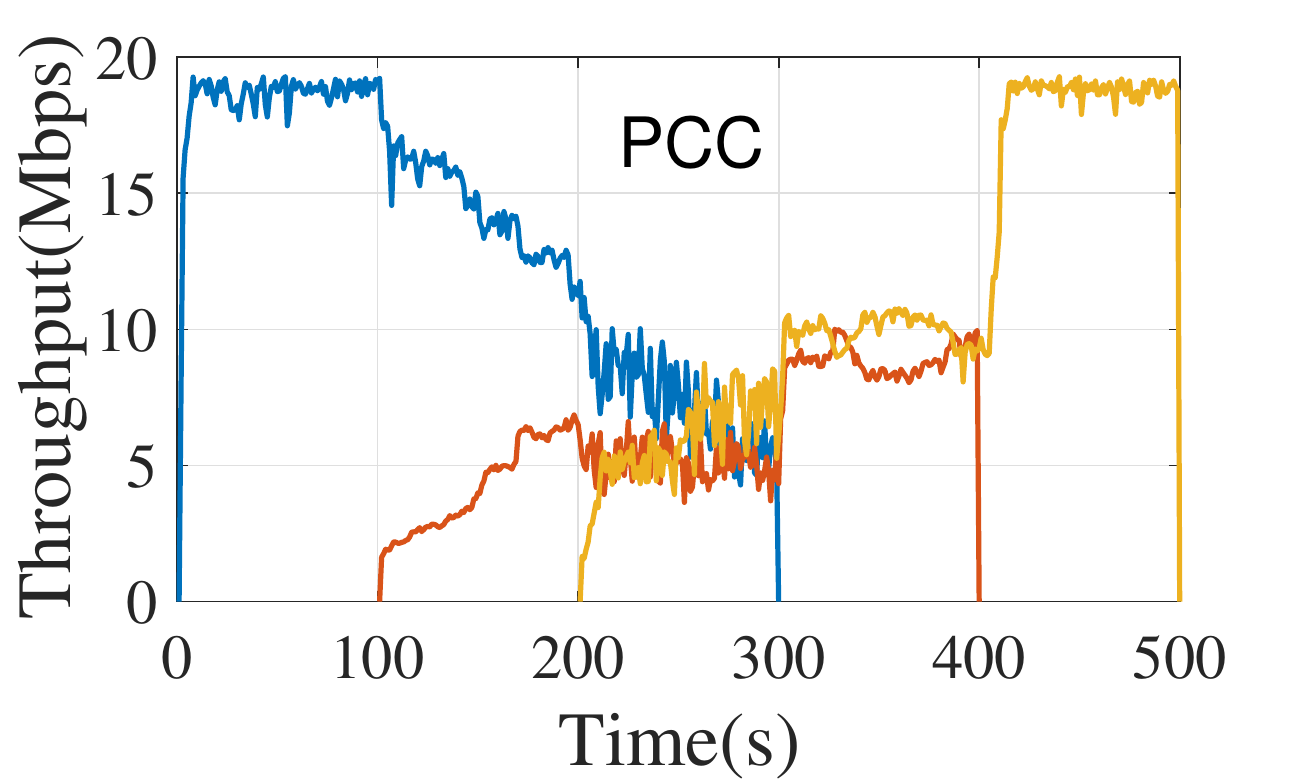}  
\end{minipage}  
}%
\subfigure[Iris Self-fairness.]{  
\begin{minipage}[t]{0.5\linewidth}  
\centering  
\includegraphics[width=1.6in]{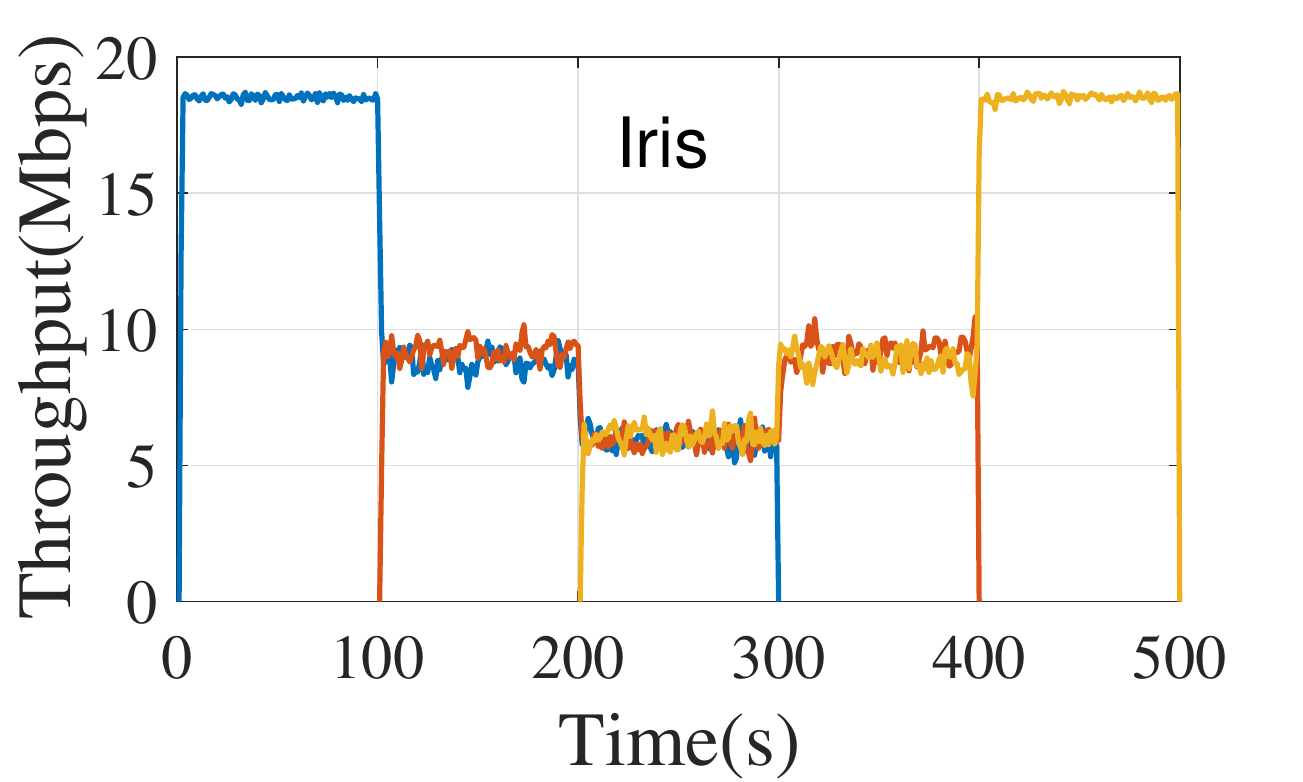}   
\label{intra-fairness-d}
\end{minipage}  
}%
\centering  
\caption{Performance of intra-protocol fairness.}  
\label{intra-fairness}
\end{figure}

\Cref{intra-fairness} shows the bandwidth allocation. Basically, Cubic, BBR, PCC and Iris all achieve a fair share of bandwidth. However, at the equilibrium point of fair bandwidth allocation, the throughput of each Cubic flow is severely jitter, which is not a stable convergence. As for BBR, the intra-protocol fairness is much better than Cubic, but there is a periodical sharp drop in throughput due to its synchronization mechanism for aligning the state of each flow. Although PCC can also share the bottleneck bandwidth fairly, its convergence time is too long, even tens of seconds, which is also confirmed in \cite{2016BBR}. In contrast, our algorithm performs better either in terms of convergence speed or stability of convergence points.

\emph{RTT fairness} \ 

Typically, the flows sharing the same bottleneck link have different propagation delay. Ideally, they should get identical bandwidth allocation, but many algorithms exhibit significant RTT unfairness, disadvantaging flows with larger RTTs. To evaluate the performance of Iris in this area, we further consider an experiment where three competing Iris flows with three different propagation delays of 50 ms, 100 ms and 150 ms share a 20 Mbps bottleneck link. 

\begin{figure}[t]
\centering
\includegraphics[width=2.8in]{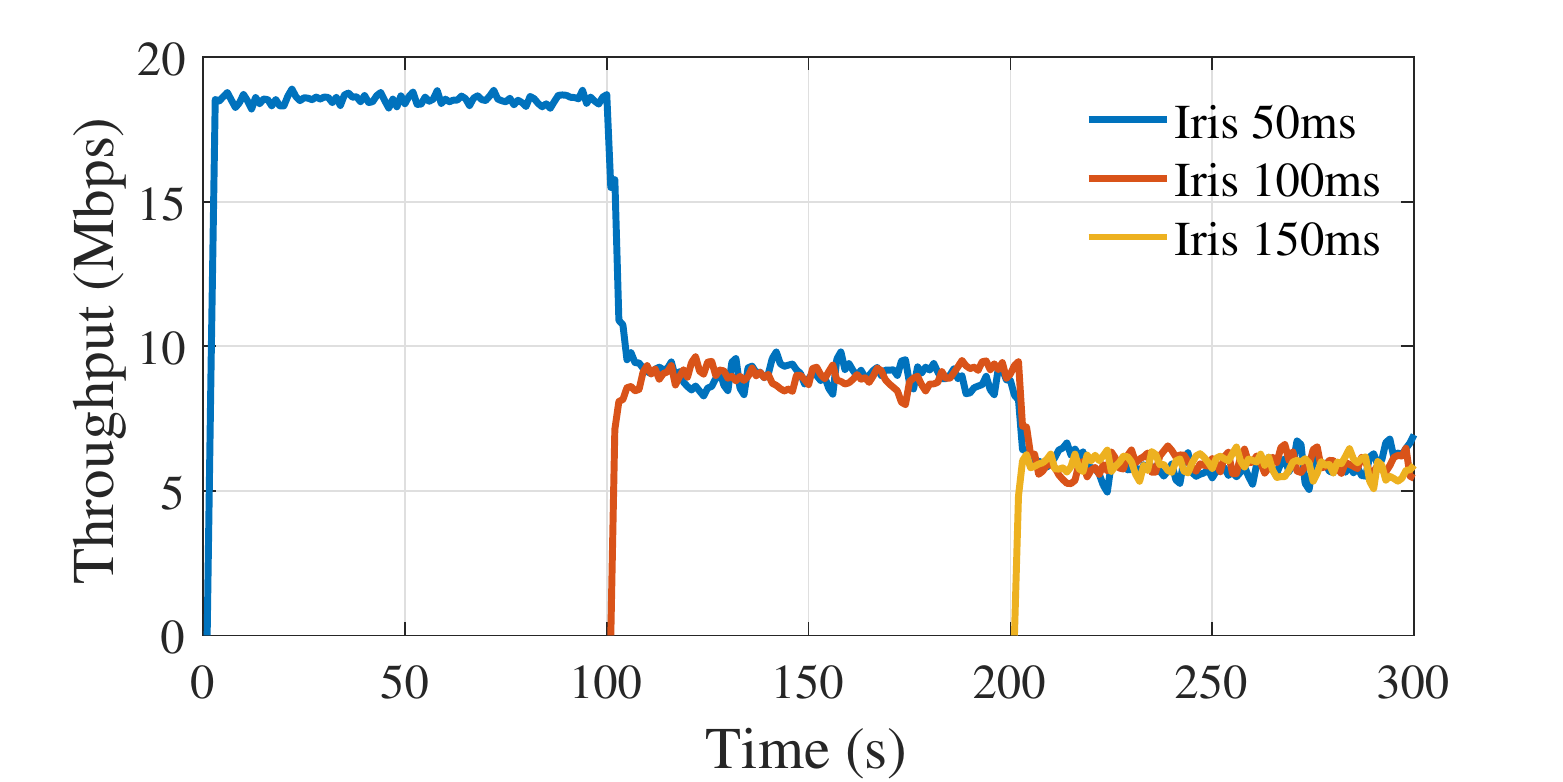}
\caption{The bandwidth allocation of three concurrent flows with different round-trip propagation delays.}
\label{fig:rtt-intra}
\end{figure}

\Cref{fig:rtt-intra} shows the temporal variation in the throughput of the three different Iris flows. Intuitively, the throughput of Iris flows is independent of the propagation delays of the flows, so \Cref{fig:rtt-intra} doesn't look very different from \Cref{intra-fairness-d}. It is due to the effect of our proposed \emph{Low-latency Fairness Model}, where the difference between $d_{i}$ and $T$ in \Cref{convergence point} can eliminate the effect of different propagation delays. It is queuing delay instead of RTT that can actually affect Iris algorithm to adjust the sending rate. \\

\emph{Fairness with TCP} \ 

Since most traffic in the Internet is still TCP-based (such as Cubic and Compound), if a congestion control algorithm is to be applied in real network environment, it is necessary to achieve a reasonable bandwidth allocation in the competition with TCP traffic. For this purpose, we use the same network as \Cref{Intra-protocol} to carry out some experiments. We run a Cubic sender as background traffic and a sender of the congestion control algorithm being evaluated. These flows are run concurrently for 20 seconds.

\begin{figure}[t]
\centering
\includegraphics[width=2.8in]{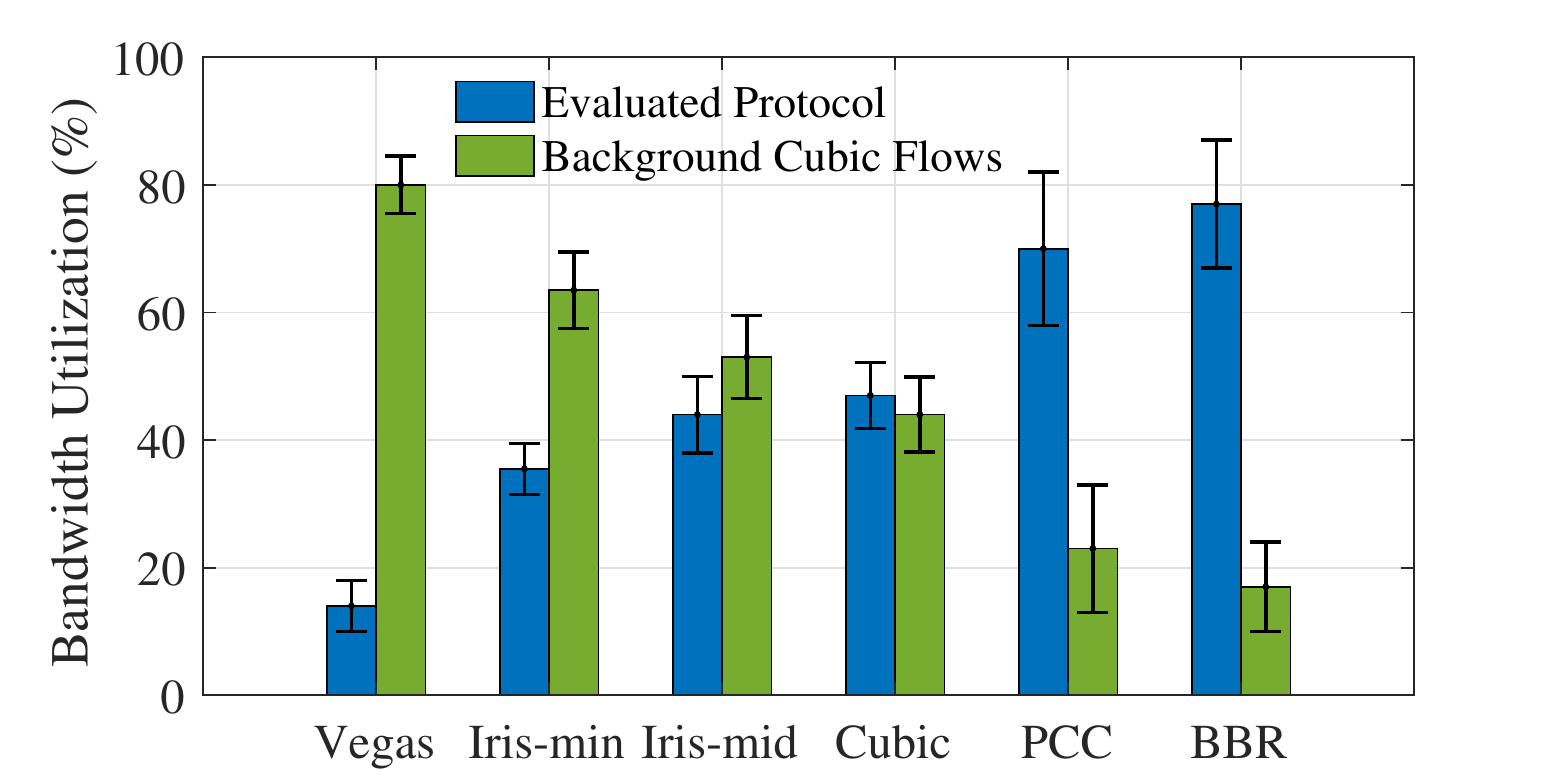}
\caption{Throughput of different algorithms versus Cubic, shown by plotting the mean and standard deviation of the bandwidth utilization.}
\label{fig:tcp-fair}
\end{figure}

\Cref{fig:tcp-fair} shows the bandwidth utilization achieved by the evaluated algorithms versus the background Cubic traffic. Intuitively, Vegas is suffering from low throughput when competing against Cubic, which is just the weakness of most delay-based algorithms. On the contrary, PCC and BBR are too aggressive, which will seriously affect other flows on the Internet. As for Iris, even if we set target delay $T$ to the minimum of RTTs (i.e. Iris-min), it can still obtain over 35\% bandwidth allocation. This is because when Cubic accumulates packets to increase RTT, the minimum value Iris sees in the time window will also increase, which will improve its convergence target. If we use the median of RTTs as the target delay $T$ (i.e. Iris-mid), the bandwidth allocation will be closer to the ground truth (i.e. Cubic v.s. Cubic) \cite{TCP-friendly}. \\

\subsubsection{Convergence Speed and Stability} \

When the network environment changes, the convergence speed and stability of the protocol determine the performance of throughput and delay. In order to measure these two aspects quantitatively, the following experiments are designed in this paper. On a link of 50 Mbps bandwidth and 50 ms RTT, for each protocol, we start 15 flows one by one at intervals of 50 seconds, which is sufficient for most of the protocols to achieve fair convergence. The convergence speed is measured by the time required for convergence, which is calculated as the time from the newer flows entry to the earliest time after which Jain's fairness index is maintained over 0.9 for at least 5 seconds. And the stability is estimated as the average standard deviation of throughput of all flows after the convergence point is reached.

\begin{figure*}[t]
	\centering
	\subfigure[Peak hours (8:00PM-9:00PM).]{
	\label{fig:Short-live}
	\includegraphics[width=0.3\textwidth]{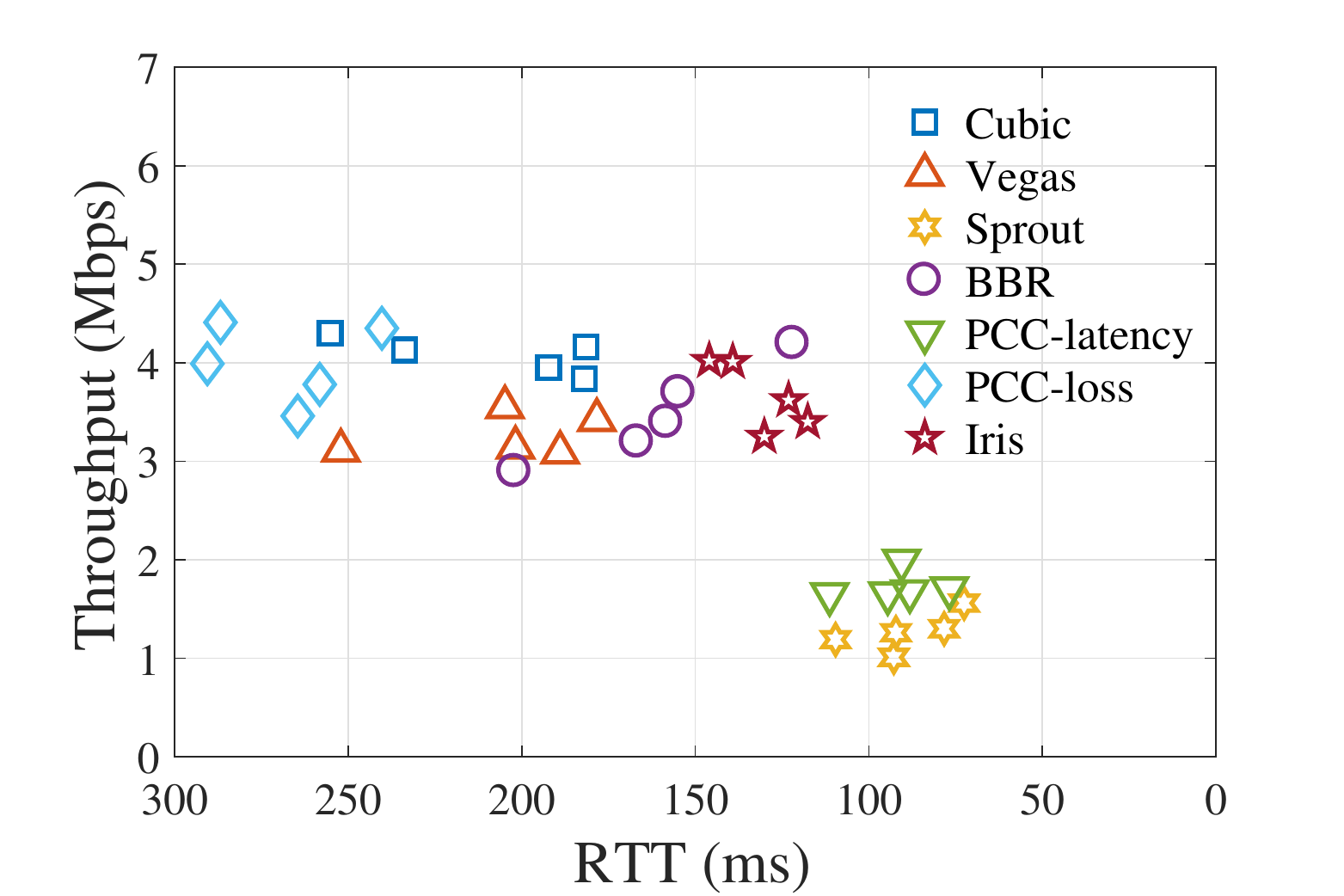}}
	\hspace{0cm}
	\subfigure[Leisure hours (10:00AM-11:00AM).]{
	\label{fig:Long-live}
	\includegraphics[width=0.3\textwidth]{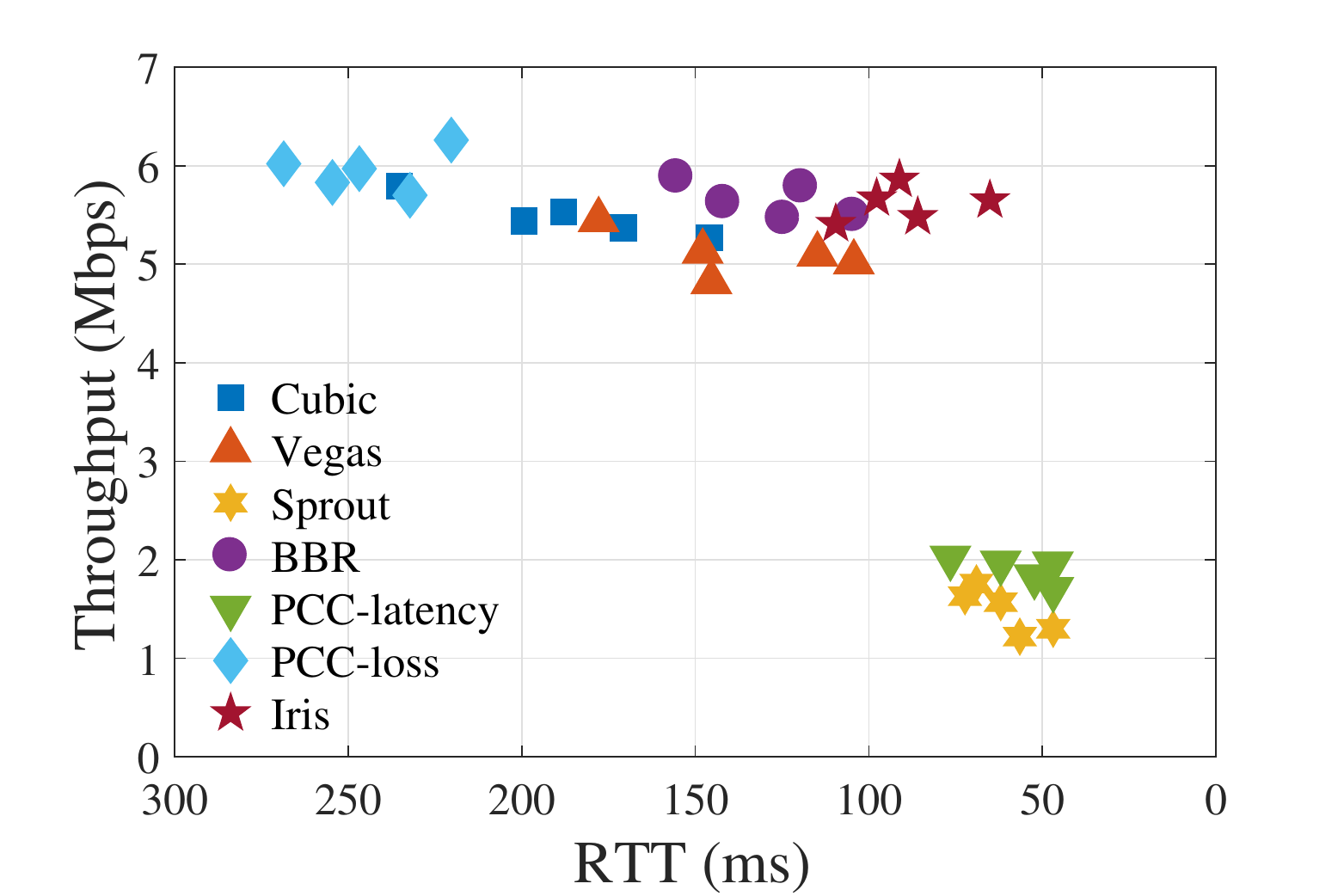}}
	\hspace{0cm}
	\subfigure[Statistical mean.]{
	\label{fig:4G-mean}
	\includegraphics[width=0.3\textwidth]{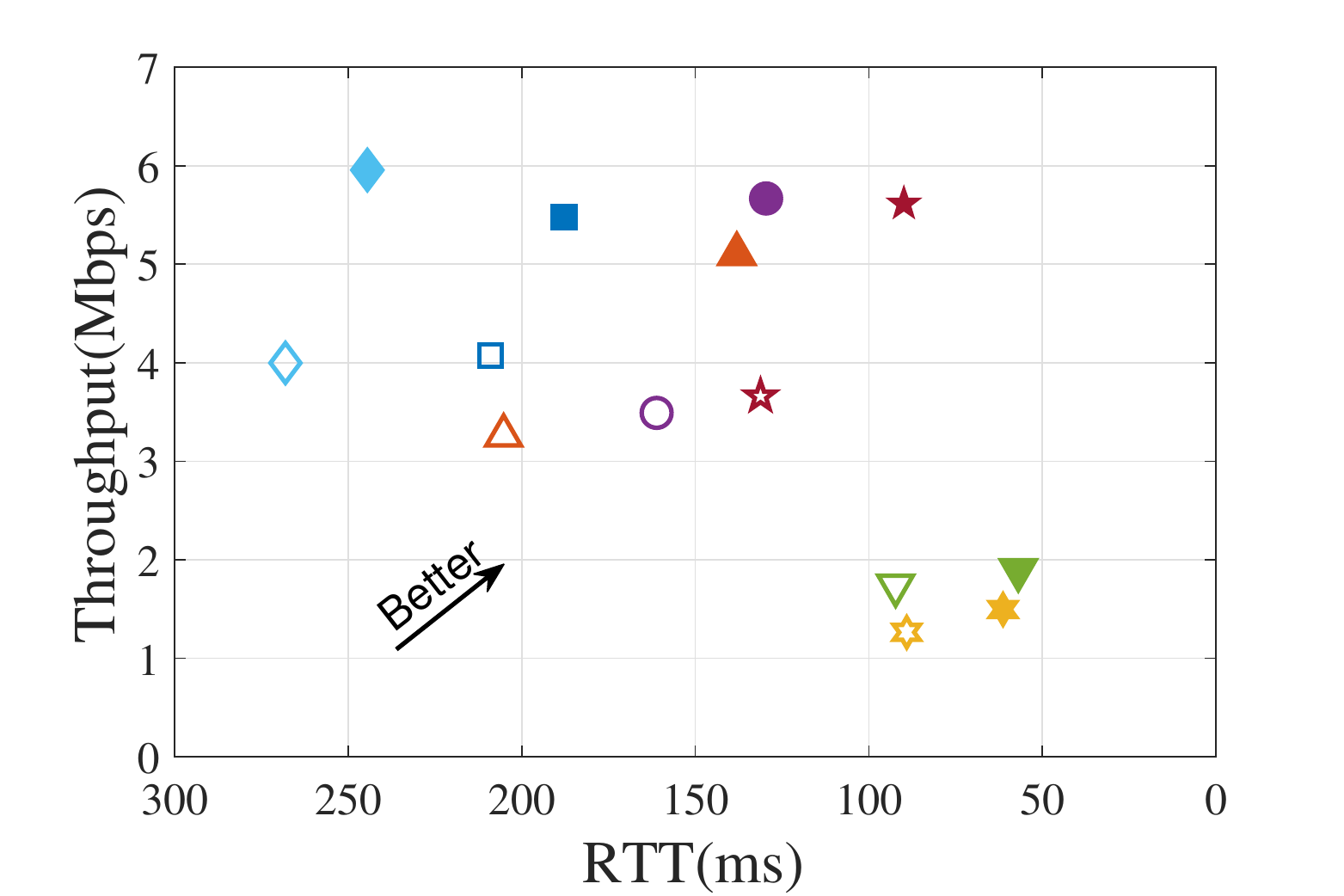}}
	\caption{Throughput and RTT in LTE networks of stationary scenario, five tests of 60s for each algorithm.}
	\label{fig:real-world-4G}
\end{figure*}

\begin{figure}[t]
\centering
\includegraphics[width=3.7in]{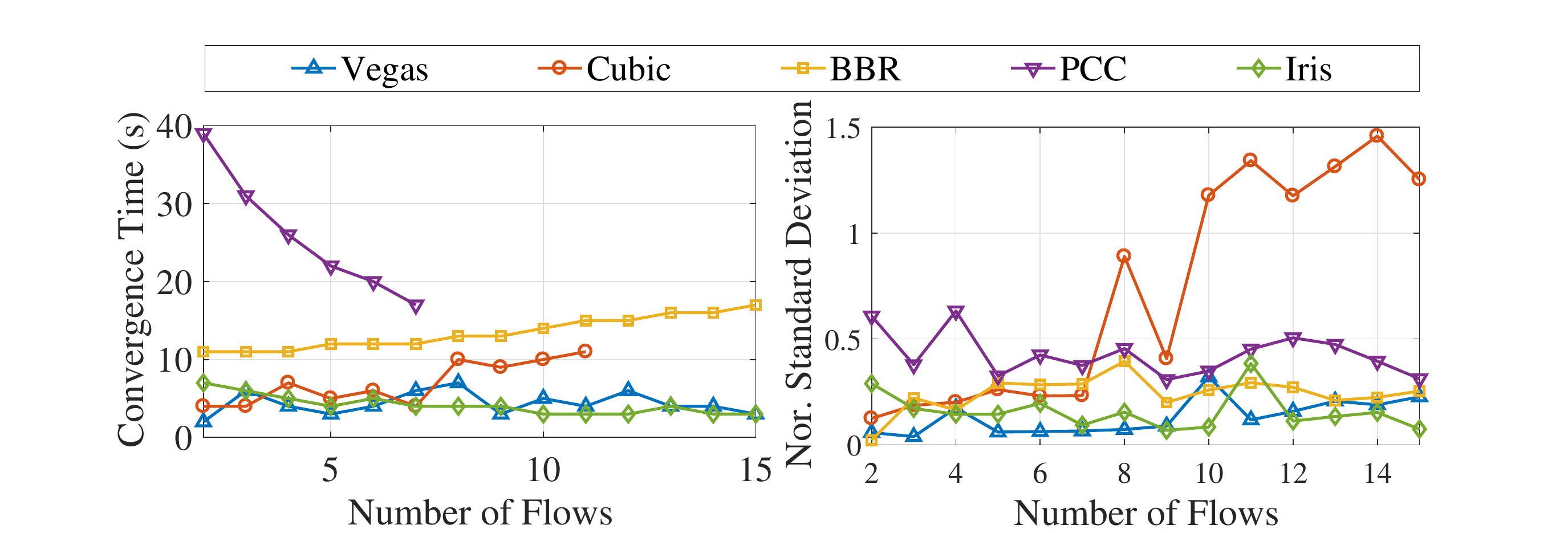}
\caption{Convergence speed and throughput stability in different concurrent numbers.}
\label{fig:StableConvergence}
\end{figure}

\Cref{fig:StableConvergence} displays the convergence time and stability as the number of flows increases.  From the view of convergence speed and the stability after convergence, our Iris and Vegas have relatively close performance, which is much better than other algorithms. It is worth noting that when the number of flows exceeds 11, Cubic can no longer achieve a fair convergence, whose bandwidth utilization will also decrease dramatically.  It is so-called ``TCP Incast'' phenomenon \cite{2009incast}. As for PCC, this threshold is even 7. Although BBR is able to achieve fair convergence, its convergence time will continue to rise with the increase of the number of flows.

\subsection{Evaluation in Real-World LTE Networks}

\subsubsection{Stationary Scenario} \
\label{Stationary Scenario}

In the real-world LTE network, an ideal congestion control algorithm can achieve high throughput and low latency at the same time. We first test the protocols in peak hours and leisure hours to evaluate the performance in stationary scenes. In addition, we evaluated PCC using loss utility function (PCC-loss) and delay utility function (PCC-latency) respectively. This evaluation was carried out in Peking University and connected to the commercial network operator, China Mobile. 

\Cref{fig:real-world-4G} displays the test results of average throughput and RTT. Intuitively, for all algorithms, the throughput in peak hours is smaller and RTT is larger. For PCC, the loss utility function makes it fill the buffer as much as possible, resulting in extremely high delay. But the utility function that tends to low delay will make it unable to get high bandwidth allocation, because its utility function can not correctly determine the direction of speed adjustment in highly variable networks. For Iris, although the throughput is not the highest, it can effectively balance throughput and delay, which is more important for real-time congestion control. \\

%

\subsubsection{Motion Scenario} \
\label{Motion Scenario}

\begin{figure}[t]
	\centering
	\subfigure[CDF of throughput.]{
	\label{fig:subway_put}
	\includegraphics[width=1.5in]{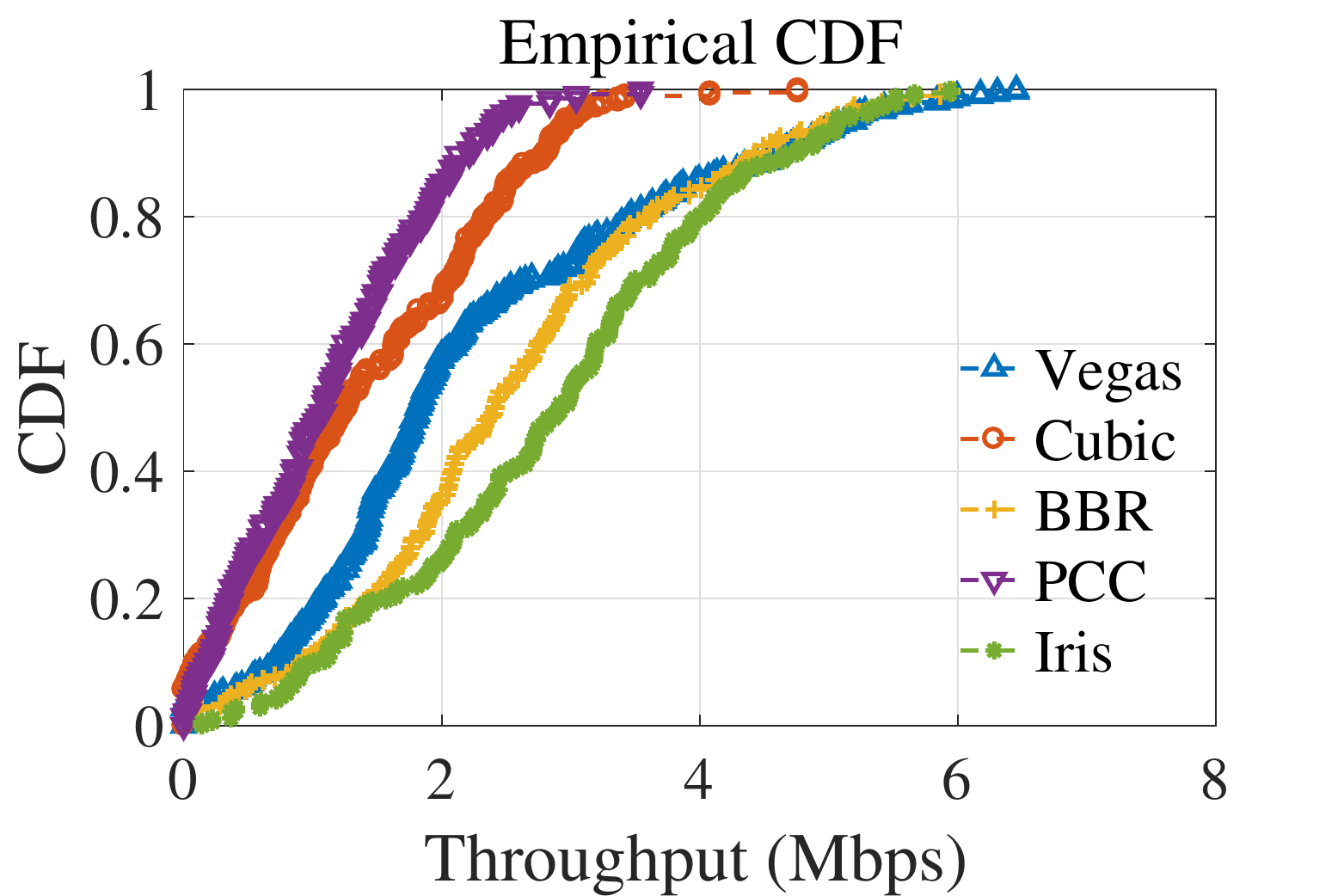}}
	\hspace{0cm}
	\subfigure[CDF of RTT.]{
	\label{fig:subway_time}
	\includegraphics[width=1.5in]{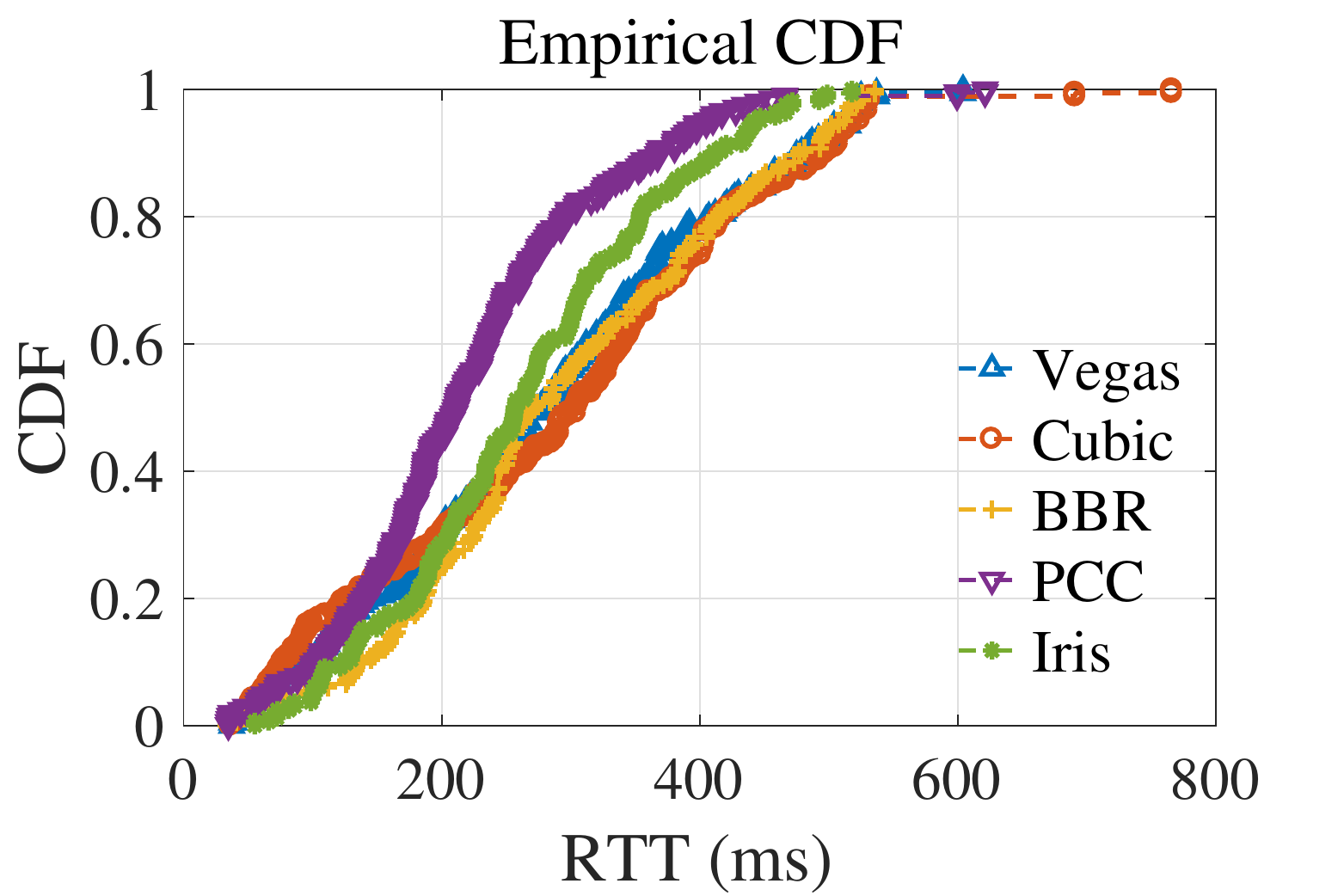}}
	\hspace{0cm}
	\caption{Performance in LTE networks of moving scenario.}
	\label{fig:subway}
\end{figure}

Furthermore, we evaluated the performance of the protocols on LTE networks in motion scenarios. Note that the subway was chosen as the test scene, mainly because there is no influence from the multi base station switching in the tunnel, so the environment is more simple. On line four of Beijing Subway, from East Gate of Peking University to Yuanmingyuan Park Station, each protocol has been tested ten times. 

\Cref{fig:subway} shows the throughput CDF of these protocols. Different with the stationary scenario, we find that the bottleneck bandwidth of the network is closely related to the location of the train when it is tested along the subway. Therefore, the measured throughput and RTT are approximately linear. In a comprehensive view, our Iris outperforms other protocols, achieving similar RTT but higher throughput. Facing more complex networks and more physical link loss in the high-speed motion scenarios, Cubic and Vegas can no longer achieve high throughput.

%
%
%
%

\begin{figure*}[t]
	\centering
	\subfigure[Peak hours (8:00PM-9:00PM).]{
	\label{fig:Short-live}
	\includegraphics[width=0.3\textwidth]{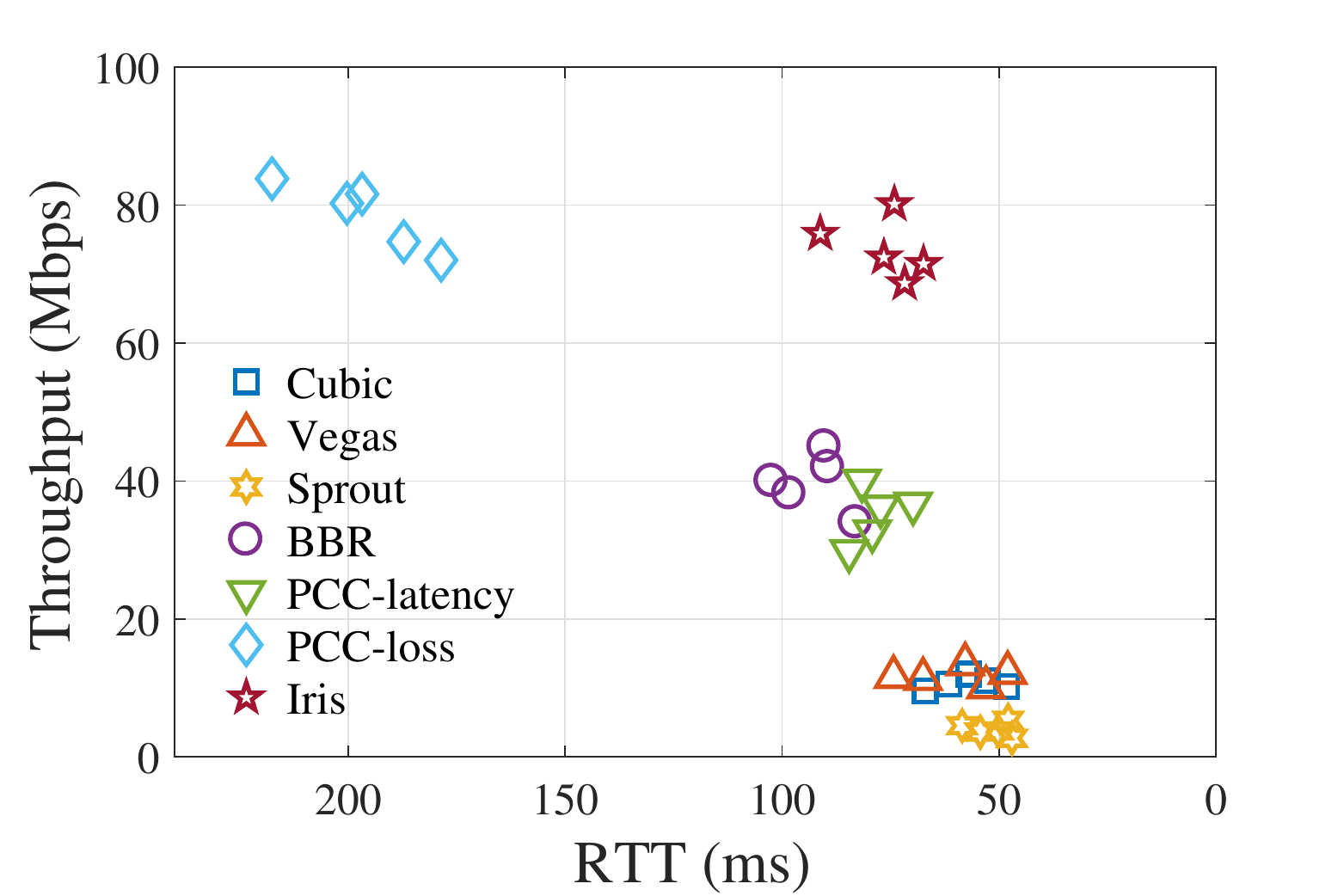}}
	\hspace{0cm}
	\subfigure[Leisure hours (10:00AM-11:00AM).]{
	\label{fig:Long-live}
	\includegraphics[width=0.3\textwidth]{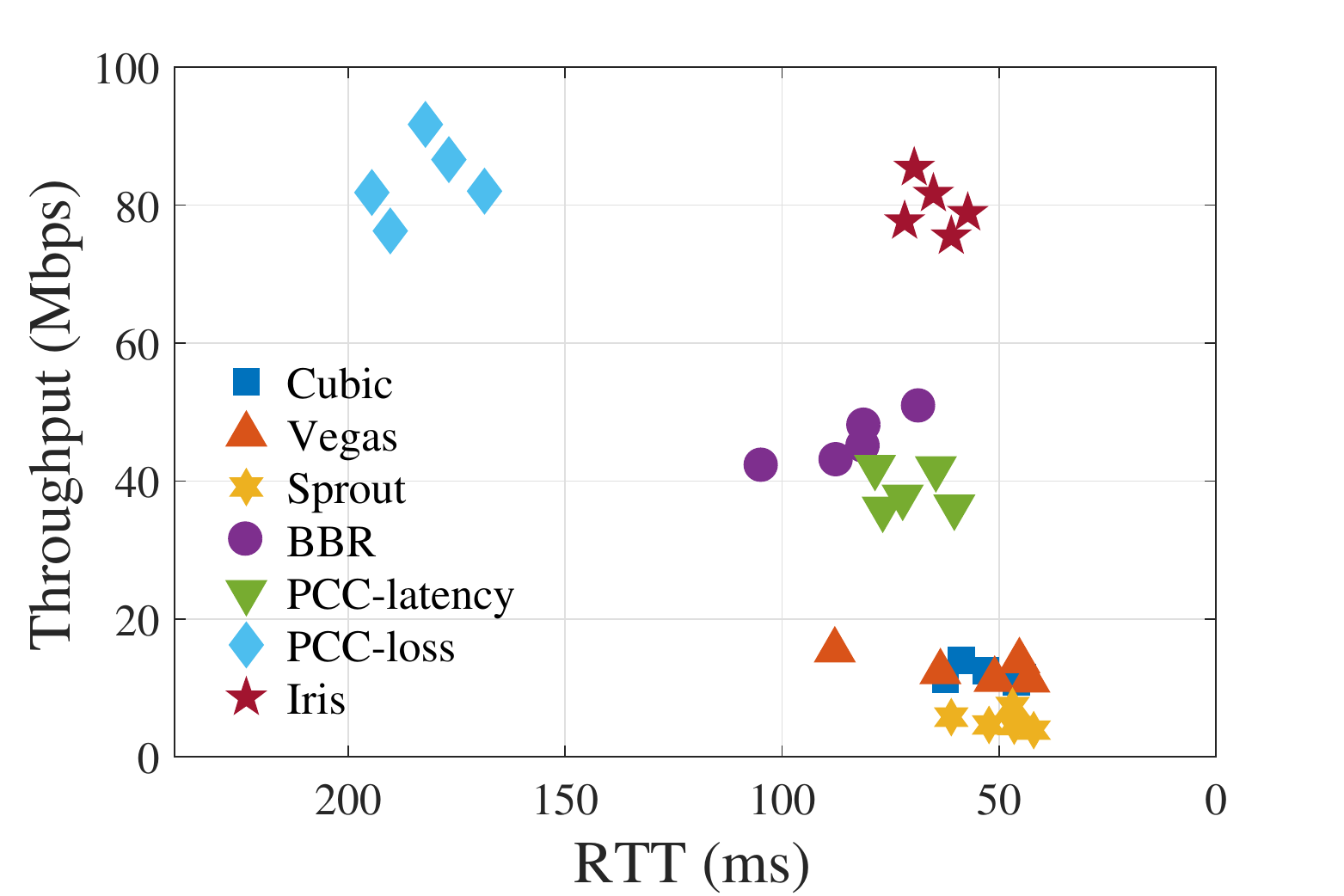}}
	\hspace{0cm}
	\subfigure[Statistical mean.]{
	\label{fig:4G-mean}
	\includegraphics[width=0.3\textwidth]{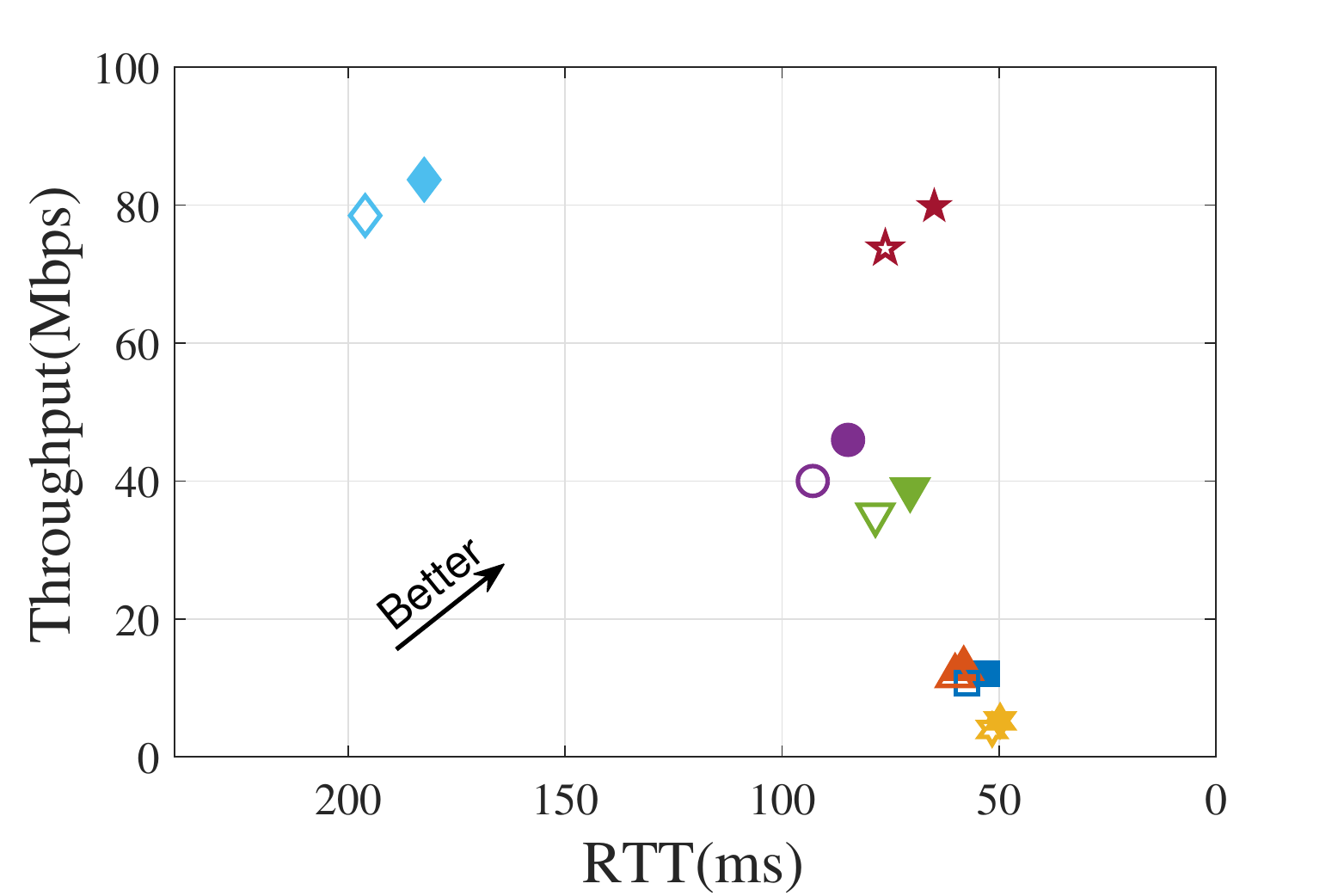}}
	\caption{Throughput vs. RTT from our host in Beijing to Ali ECS in HongKong, five tests of 60s for each algorithm.}
	\label{fig:real-world-host}
\end{figure*}

\begin{figure*}[htbp]
	\centering
	\subfigure[Using Cubic congestion control.]{
	\label{fig:quic_cubic}
	\includegraphics[width=0.31\textwidth]{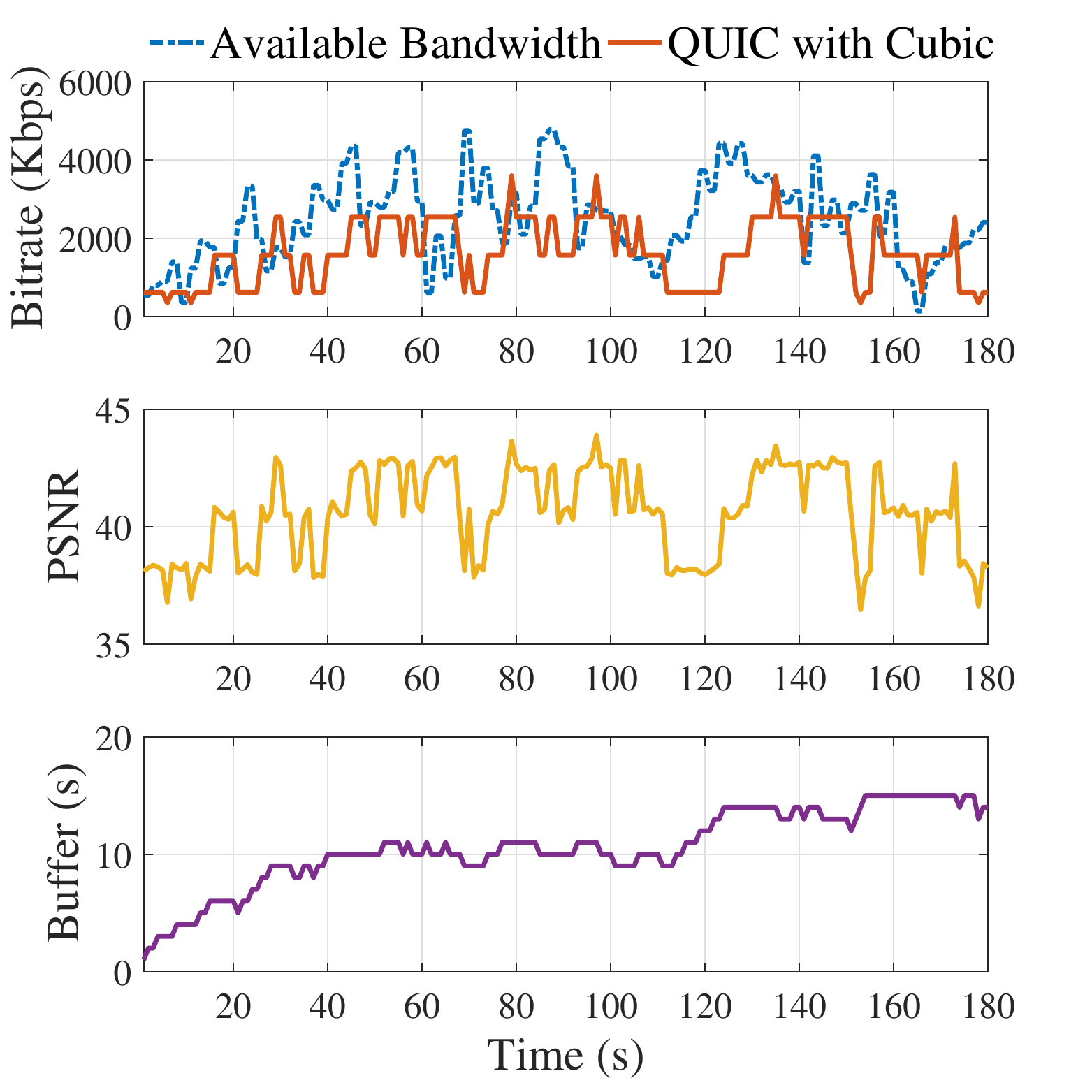}}
	\hspace{0cm}
	\subfigure[Using BBR congestion control.]{
	\label{fig:quic_bbr}
	\includegraphics[width=0.31\textwidth]{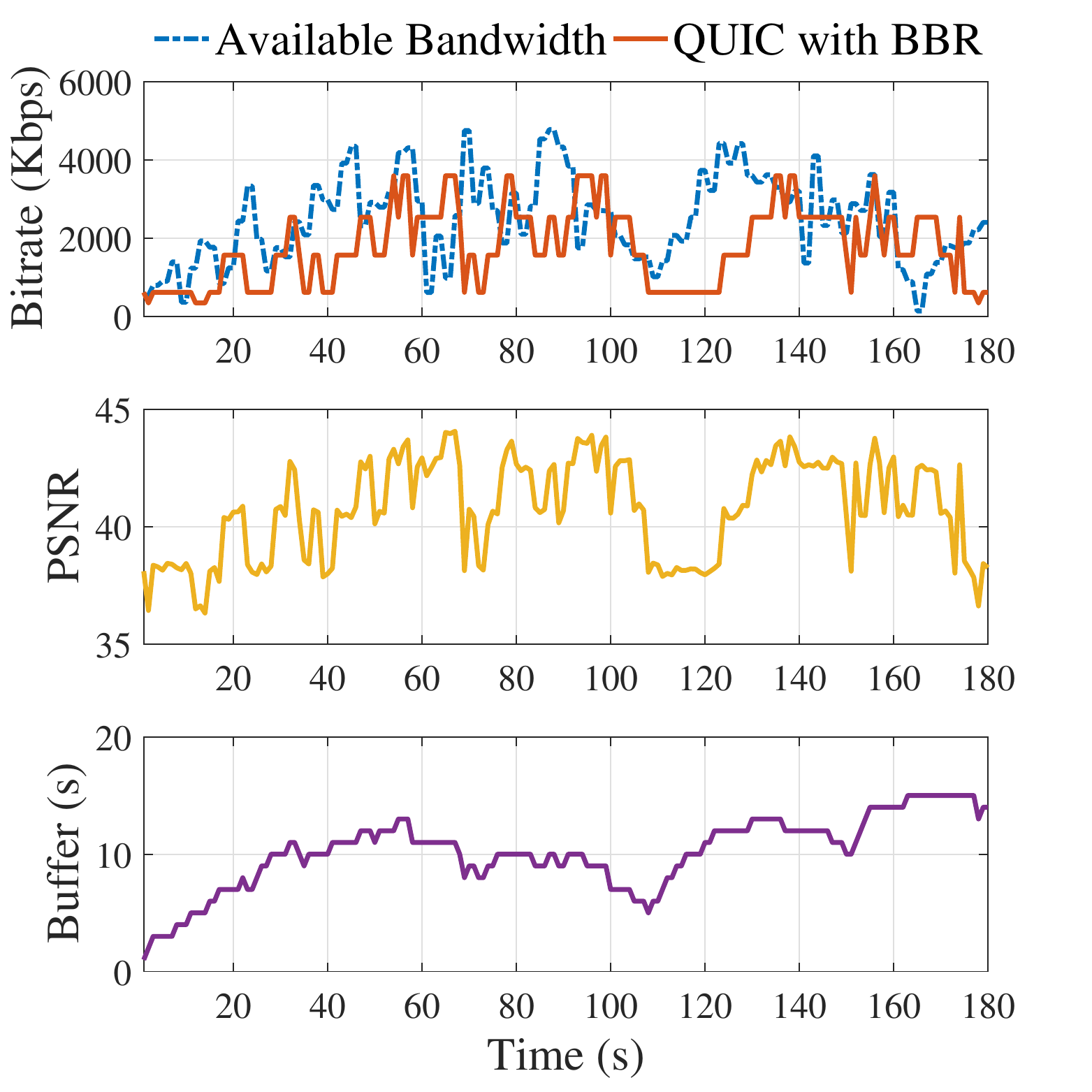}}
	\hspace{0cm}
	\subfigure[Using Iris congestion control.]{
	\label{fig:quic_iris}
	\includegraphics[width=0.31\textwidth]{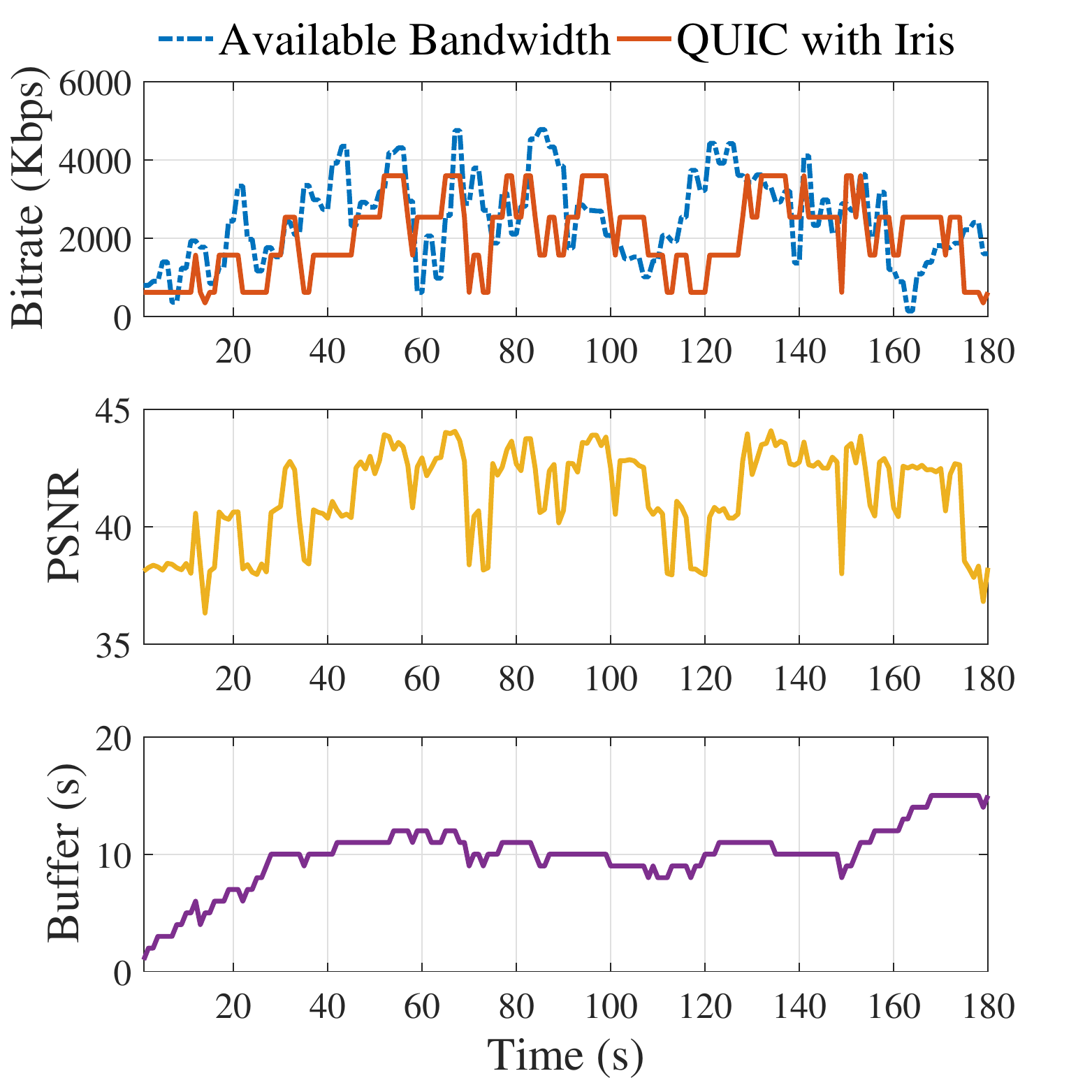}}
	\caption{The bitrate, PSNR and buffer occupancy during video playback when using different congestion control algorithms.}
	\label{fig:http_quic}
\end{figure*}

\subsection{Evaluation in Intercontinental Networks}

To evaluate the protocols in the intercontinental networks, we employ Ali ECS to establish four nodes in Beijing, HongKong, Singapore and America respectively. Our host in Beijing is used as a sender while these nodes are regarded as receivers, forming four links.

\begin{table}[t]
\caption{Average throughput (Mbps) and RTT (ms) vs. Iris on the link from Beijing to the Ali ECS Beijing, HongKong, Singapore and America.}
\centering
\scalebox{0.76}{
\begin{tabular}{ccccccccc}
\toprule[1.1pt]
Link &\multicolumn{2}{c}{BJ $\rightarrow$ BJ}&\multicolumn{2}{c}{BJ $\rightarrow$ HK}&\multicolumn{2}{c}{BJ $\rightarrow$ SG}&\multicolumn{2}{c}{BJ $\rightarrow$ USA}\\
\midrule
Protocol&Rate&RTT&Rate&RTT&Rate&RTT&Rate&RTT\\
Iris& 83.6 & 28 & 76.2 & 66 & 62.2 & 144 & 38.4 & 201 \\
Vegas &$0.52\times$&$0.78\times$&$0.19\times$&  $0.79\times$&$0.22\times$&$0.93\times$&$0.21\times$&$0.93\times$\\
Cubic &$0.48\times$&$0.80\times$& $0.20\times$&  $0.74\times$ &$0.27\times$&$0.93\times$&$0.19\times$&$0.93\times$\\
BBR&$0.73\times$&$1.07\times$&$0.56\times$&$1.11\times$& $0.39\times$&$0.99\times$&$0.37\times$&$0.97\times$\\
Sprout &$0.05\times$ &$0.36\times$&$0.06\times$&$0.68\times$& $0.06\times$&$0.90\times$&$0.06\times$&$0.91\times$\\
PCC-Loss &$1.08\times$ &$5.85\times$&$1.07\times$&$2.85\times$& $0.90\times$&$1.31\times$&$1.06\times$&$1.17\times$\\
PCC-Latency &$0.69\times$ &$0.96\times$&$0.49\times$&$1.08\times$& $0.41\times$&$1.01\times$&$0.32\times$&$0.94\times$\\
\bottomrule[1.1pt]
\end{tabular}}
\label{host-ECS}
\end{table}

Taking the link from the host in Beijing to Ali ECS in HongKong as an example, \Cref{fig:real-world-host} shows  throughput and RTT of the protocols, in peak hours and leisure hours respectively.  Intuitively, our Iris algorithm outperforms others. Different from the results in \Cref{fig:real-world-4G}, the bandwidth utilization of Cubic and Vegas has decreased significantly, while the throughput of PCC has increased a lot, mainly because the complex background traffic often brings packet loss, but delay jitter is more gentle than LTE network. Our Iris is more competitive in the competition scenario, so it can still maintain high throughput, no matter at peak or leisure hours. In addition, we use the same method to evaluate the algorithms on the other three links. \Cref{host-ECS} displays the average throughput and RTT of these protocols vs. Iris, which further proves the excellence of Iris.

\subsection{Performance in HTTP over QUIC}

To evaluate the gain of Iris congestion control algorithm for Internet video transmission, we built an HTTP live streaming system with QUIC server \cite{quic_code} and MPEG-DASH \emph{dash.js} \cite{dashjs}. Notice that the adaptive bitrate (ABR) algorithm we used in \emph{dash.js} is a simple method based on bandwidth estimation.

As depicted in \Cref{fig:http_quic}, we compare our Iris with Cubic and BBR under a real Internet bandwidth trace (about 180 seconds), where both the long-term shifts and short-term fluctuations of bandwidth can be observed. Besides, the round-trip propagation delay is set to 50ms and random loss is 1\%. Due to the conservative bandwidth-based adaption logic, there is no stalling phenomenon in the process of video playback. However, because of the low bandwidth utilization of Cubic congestion control in this case, the client rarely requests the highest bitrate video segments, as shown in \Cref{fig:http_quic}. And the client with BBR algorithm has higher instability in bitrate, compared with Iris. 

\begin{figure}[tbp]
\centering
\includegraphics[width=3.2in]{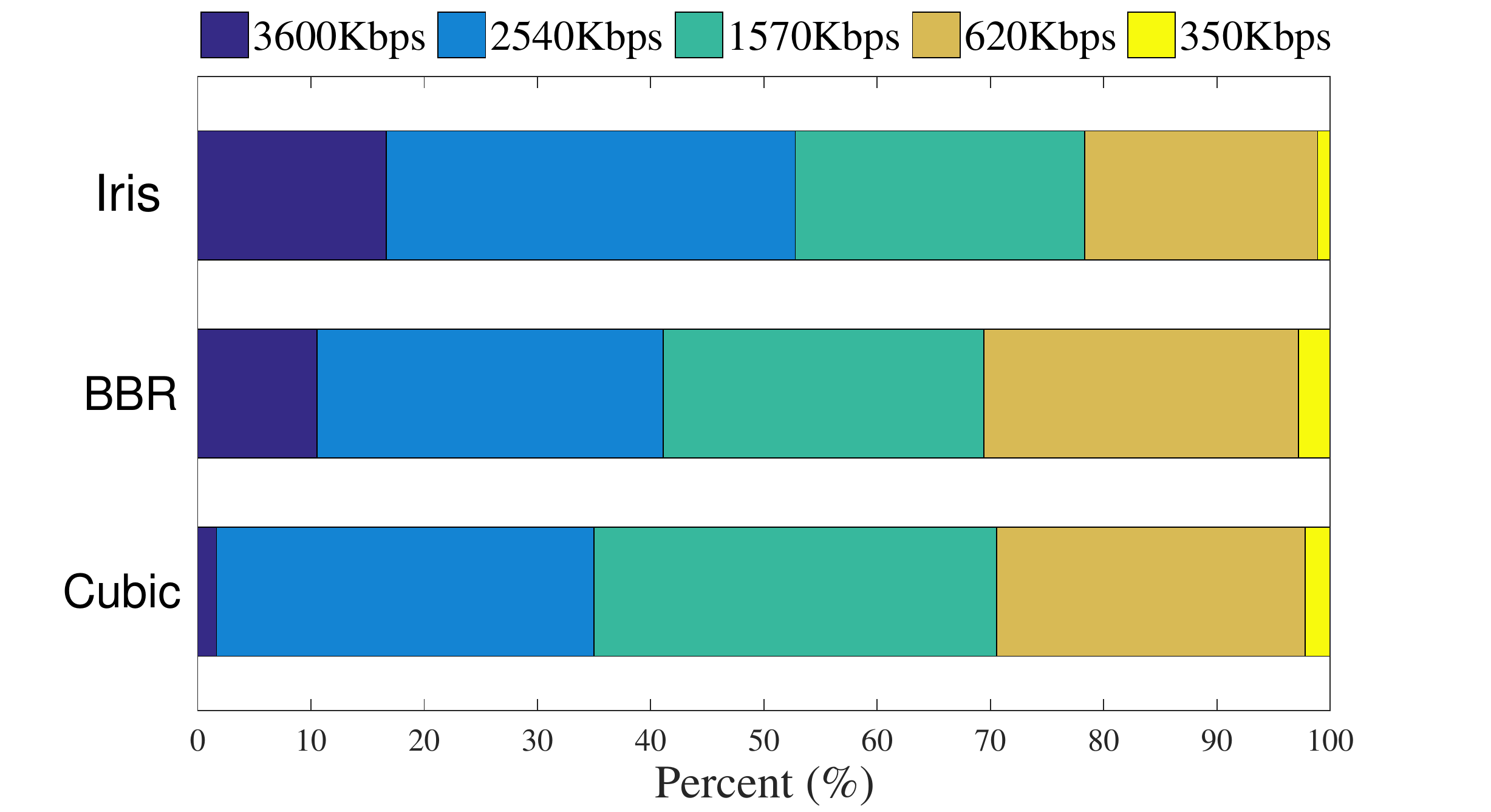}
\caption{The proportion of video bitrate level.}
\label{fig:http_compare}
\end{figure}

In order to show the difference more clearly, we summarize the bitrate proportion of the requested video segments, as depicted in \Cref{fig:http_compare}. It demonstrates that more than 50\% of the video segments requested by Iris are 3600Kbps or 2540Kbps, while this percentage is only about 35\% for Cubic. Quantitatively, as shown in \Cref{bitrate_psnr} , Iris achieves a 25\% bitrate improvement over Cubic and 15\% over BBR. And the average PSNR of Iris is also the highest, compared against Cubic and BBR. These results indicate that replacing the congestion control module in QUIC with our proposed Iris is able to achieve higher user experience quality.

\newcommand{\tabincell}[2]{\begin{tabular}{@{}#1@{}}#2\end{tabular}}
\begin{table}[tbp]
\caption{Performance under the real Internet trace}
\centering
\scalebox{1}{
\begin{tabular}{ccc}
\toprule[1.1pt]
Protocol & \tabincell{c}{Average\\ PSNR (dB)}  & \tabincell{c}{Average\\ bitrate (Kbps)} \\
\midrule
Cubic & 40.536 & 1624.4\\
BBR & 40.712 & 1763.5\\
Iris & 41.27 & 2033.8\\
\bottomrule[1.1pt]
\end{tabular}}
\label{bitrate_psnr}
\end{table}

\section{Conclusions}
\label{CONCLUSIONS}

In this paper, we have designed Iris, an end-to-end statistical learning based congestion control algorithm. The key ideas of Iris are keeping a small and fixed \emph{queue load} in networks, and adaptively adjusting sending rate based on a linear-regression learning model. By forcing all flows to maintain the same queue load, a fair share of bandwidth and low latency are both achieved. The rate can be periodically updated by online regression learning, which avoids hardwired rate adjustment to better adapt to dynamically changing networks. Extensive experiments are carried out to evaluate the performance of Iris. It shows that, in various network environments, Iris is able to achieve high bandwidth utilization and low latency at the same time, and outperforms most existing algorithms. It also improves Internet video services in the application layer, increasing the bitrate by 25\% compared against Cubic, under the real network trace.



\section*{Acknowledgment}

The authors would like to thank all reviewers for providing many constructive suggestions which will significantly improve the presentation of this paper.

\ifCLASSOPTIONcaptionsoff
  \newpage
\fi



%

\bibliographystyle{IEEEtran}
\bibliography{reference}

\end{document}